\pdfoutput=1
\documentclass[12pt,a4paper]{article}
\usepackage{a4wide,hyperref}
\usepackage{subcaption}
\usepackage[english]{babel}
\usepackage{amsmath}   
\usepackage{graphicx}  
\usepackage{dcolumn}   
\usepackage{bm}        
\usepackage{amssymb}   
\usepackage{epsfig}
\usepackage{epstopdf}
\usepackage[utf8]{inputenc}
\usepackage{mdframed}
\usepackage{amsmath}
\usepackage{wasysym}
\usepackage{tabu}
\usepackage[toc]{appendix}
\usepackage{color,soul} 
\usepackage{datetime}
\usepackage{IEEEtrantools}
\usepackage{tikz}
\usetikzlibrary{calc, 3d}
\usepackage{physics}
\usepackage{enumerate}
\usepackage{asymptote}
\usepackage{pifont}

\newcommand{\be}{\begin{equation}}
\newcommand{\ee}{\end{equation}}
\newcommand{\bea}{\begin{eqnarray}}
\newcommand{\eea}{\end{eqnarray}}
\newcommand{\bean}{\begin{eqnarray*}}
\newcommand{\eean}{\end{eqnarray*}}

\newcommand{\eps}{\epsilon}

\begin{document}

\begin{titlepage}

\numberwithin{equation}{section}
\begin{flushright}
\small

\normalsize
\end{flushright}
\vspace{0.8 cm}

\begin{center}

\hspace*{-1cm}\mbox{\LARGE \textbf{Gravitational Waves, Holography,}}\\ \vspace{0.5cm}
{\LARGE \textbf{and Black Hole Microstates}}

\medskip

\vspace{1.2 cm} {\large Vasil Dimitrov$^1$, Tom Lemmens$^{1}$, Daniel R. Mayerson$^2$, Vincent S. Min$^1$, Bert Vercnocke$^1$}\\

\vspace{1cm} {$^1$ Institute for Theoretical Physics, KU Leuven,\\
 Celestijnenlaan 200D, B-3001 Leuven, Belgium}

\vspace{0.5cm} {$^2$ Universit\'e Paris-Saclay, CNRS, CEA,\\ Institut de physique th\'eorique,\\ 91191, Gif-sur-Yvette, France.}

\vspace{0.5cm}

\vspace{.5cm}
vasko.dimitrov @ kuleuven.be, tom.lemmens @ kuleuven.be, daniel.mayerson @ ipht.fr, \\
vincent.min @ kuleuven.be,
bert.vercnocke @ kuleuven.be\\

\vspace{2cm}

\textbf{Abstract}
\end{center}

\noindent Gravitational wave observations of the near-horizon region of black holes lend insight into the quantum nature of gravity. In particular, gravitational wave echoes have been identified as a potential signature of quantum gravity-inspired structure near the horizon. In this paper, we connect such observables to the language of black hole microstates in string theory and holography. To that end, we propose a toy model describing the AdS$_3$ near-horizon region of five-dimensional black holes, inspired by earlier work of Solodukhin. This model captures key features of recently constructed microstate geometries, and allows us to make three observations. First, we relate the language of AdS/CFT, in particular the holographic retarded two-point correlator, to effective parameters modeling the structure that are used in flat space gravitational wave literature. Second, we find that for a typical microstate, the `cap' of the microstructure is exponentially close to the horizon, making it an effective sub-Planckian correction to the black hole geometry, although the microstate geometry itself is classical. Third, using a microcanonical ensemble average over geometries, we find support for the claim that the gravitational wave echo amplitude in a typical quantum microstate of the black hole is exponentially suppressed by the black hole entropy.

\end{titlepage}

\newpage

\setcounter{tocdepth}{2}
\tableofcontents


\section{Introduction}\label{sec:intro}
Gravitational wave echoes in the late ringdown signal have been at the heart of searches for quantum structure near black hole horizons since the work of \cite{Cardoso2016(2),Cardoso2016} and the claim that they were observed in the first LIGO run \cite{Abedi:2016hgu}, see \cite{Cardoso2017,Cardoso2017(3),Cardoso:2017cqb,Conklin:2017lwb,Mark:2017dnq,Bueno:2017hyj,Price:2017cjr,Wang:2018mlp,Wang:2019szm,Maggio:2019zyv}.
Although a more detailed data analysis so far has tempered expectations and has shown no conclusive evidence of such effects in gravitational wave (GW) observations \cite{Ashton:2016xff,Abedi:2017isz,Westerweck:2017hus,Abedi:2018pst,Lo:2018sep,Nielsen:2018lkf,Testa:2018bzd,Tsang:2018uie,Abedi:2018npz,Abedi:2020sgg,Uchikata:2019frs}, echoes remain a prime target for future gravitational wave searches.

For echoes to appear in the late-time ringdown signal, one needs a change in the absorbing boundary conditions of the wave equation at the location of the event horizon. Echoes follow naturally from waves being temporarily trapped in the cavity that forms between the light-ring and the reflective structure, while the quasinormal frequencies of the structure are determined by the inverse travel time in that cavity and the behaviour  near the light ring.  
Potentially detectable effects extend beyond echoes, and include tidal effects such as tidal deformabilities, tidal heating and inspiral resonances. We refer to the reviews \cite{Carballo-Rubio:2018jzw,Cardoso:2019rvt} for more references.

How well do we expect gravitational waves to probe the near-horizon structure of black holes, from a fundamental theory point of view?   One can (and should) wonder how sensible these reflections, and effects associated with them, can be in a quantum theory of gravity. Since we expect the black hole to be described by a microstate or ensemble of microstates, it is crucial to understand the microscopic and statistical physics underlying the thermodynamic description of black holes in semi-classical gravity.

In string theory, certain microstates can be described as geometries with structure at the horizon scale. Those are the `microstate geometries' that are built within the fuzzball program \cite{Mathur:2005zp,Mathur:2009hf,Bena:2013dka,Bena:2007kg,Shigemori:2020yuo} and they indeed share similar reflective properties with other compact objects. The scattering of scalar waves, including the appearance of echoes, was studied in certain microstate geometries in gravity and their holographically dual CFT state in the early days of the fuzzball proposal over 15 years ago \cite{Giusto:2004id,Giusto:2004ip,Giusto:2004kj,Jejjala:2005yu,Cardoso:2005gj}; trapping, the 
slow decay and quasinormal frequencies were discussed in \cite{Chowdhury:2007jx,Chakrabarty:2015foa,Eperon:2016cdd, Eperon:2017bwq, Keir:2018hnv,Chakrabarty:2019ujg}. This was recently extended to more intricate ``superstratum'' solutions in \cite{Bena:2019azk,Bena:2020yii}. 

However, all of the solutions in question are smooth, classical geometries and thus correspond to coherent states. Scattering of waves off a \emph{typical} microstate of a black hole may behave quite differently; studying this would require methods that go beyond classical geometries.
Quantum effects might well wash out the geometric structure near the location of the horizon, leading to corrections to the semi-classical gravitational observables that are exponentially suppressed in the entropy as $e^{-S_{\rm BH}/2}$ \cite{Balasubramanian:2005qu,Balasubramanian:2007qv,Balasubramanian:2008da}. This would make a typical microstate all but indistinguishable from the black hole \cite{Raju:2018xue}; see for example the construction of operators for asymptotically Anti-de Sitter black holes in \cite{Papadodimas:2012aq,Papadodimas:2013jku,Papadodimas:2013wnh,Papadodimas:2013kwa,Papadodimas:2015jra,Banerjee:2016mhh,deBoer:2019kyr}. This cast doubt on whether gravitational waves can actually successfully probe and capture information on the particular black hole microstate or fuzzball describing an astrophysical black hole.	

In this paper, we want to provide a step towards making these statements and questions quantitative. We revisit the scattering of scalar waves, and focus on how a classical, asymptotic observer would interact with the microstates of the black hole offered to us by string theory. Such string theory microstates are best understood through the notion of holography; the degrees of freedom of the near-horizon region of black hole spacetimes are described by a spacetime that is asymptotically Anti-de Sitter (AdS), which is holographically dual to a conformal field theory (CFT) living on its boundary. 
At present, it is not possible to look into this for four-dimensional Kerr black holes, as the microstates of that system have not been mapped out in detail (although see \cite{Bena:2012wc,Bena:2015pua,Heidmann:2018mtx} for recent work in this direction). Therefore, we keep our focus on arguably the most studied black hole in string theory: the five-dimensional asymptotically flat three-charge black hole. Near the horizon, the fibration of one of the internal dimensions over five-dimensional spacetime leads to an $\text{AdS}_3 \times S^3$ region, dual to the two-dimensional D1-D5 orbifold CFT. This near-horizon limit, obtained after a specific decoupling limit, is described by the three-dimensional BTZ black hole; so it is more appropriate to speak of the  $\text{BTZ}_3\times S^3$ decoupling region. 
	
For concreteness, we consider the scattering of massless scalar fields as a first step towards richer probes such as GWs (in addition, GW perturbations also have a massless scalar component).  To highlight the relation between asymptotically flat observables and the CFT data, we wil consider a massless scalar field coming in from (flat) infinity in a five-dimensional three-charge black hole background, and descending into the decoupling region. We will describe how the asymptotic behaviour is related to the holographically dual field theory.  This link between holography in the BTZ decoupling limit and flat space has been pointed out before by using matched asymptotic expansions between the decoupling region and the asymptotically flat region \cite{Giusto:2004id,Giusto:2004ip,Giusto:2004kj,Jejjala:2005yu,Cardoso:2005gj}, and very recently using a \emph{hybrid WKB method} for microstates that do not have a separable wave equation outside the decoupling region \cite{Bena:2019azk,Bena:2020yii}.

We expand on those ideas in a controlled toy-model setup. It is known that the quasinormal modes (QNMs) of the scalar wave equation in the BTZ region are identical to the poles of the frequency space retarded two-point correlator in the two-dimensional CFT \cite{Birmingham:2001pj,son_minkowski-space_2002}. Therefore, most of our attention will go to discussing the solutions of the wave equation in the decoupling limit, which can be posed as a Schr\"odinger problem, and relating these solutions to the two-point correlator. For reasons of computation, we chose not to do this in a  specific microstate solution. We study instead a rotating variation of the asymptotically AdS wormhole solution introduced by Solodukhin in \cite{solodukhin_restoring_2005}, whose four-dimensional variation by Damour and Solodukhin had one of the first discussions of GW echoes \cite{damour_wormholes_2007,Bueno:2017hyj}. 
The benefit of this simple solution is that we have one controllable parameter $\lambda$ which determines the position of the throat and hence the deviation from a black hole, while we still capture some of the main features expected from classical microstate geometries in $\text{AdS}_3 \times S^3$. 

Using our model, we deliver three key messages that we summarize in section \ref{sec:results}; the structure of the rest of the paper is outlined in section \ref{sec:structure}.

\subsection{Results}\label{sec:results}

\paragraph{Toy models can capture many qualitative features of black hole microstates.} Black hole microstates in string theory are generally very complicated objects, and calculating correlators in them is challenging \cite{Bena:2019azk}. However, toy models such as the Solodukhin wormhole can capture qualitative features of these microstates, facilitating analysis of correlators. For example, we show that the structure of the scalar two-point correlator in the extremal BTZ Solodukhin wormhole mimicks very well the structure of the correlator in the (much more complicated) superstratum geometry, using the `hybrid' WKB method developed in \cite{Bena:2019azk}. 

Using a matched asymptotic expansion, we find the precise way in which the flat space QNMs, and hence the time-behaviour, are related to the solution of the scalar wave equation in the decoupling limit. The real part of the QNMs is given to excellent approximation by the real part of the QNMs in the decoupling region, ${\rm Re }\, \omega \sim (t_{\rm travel})^{-1}$, where $t_{\rm travel}$ is the travel time from the AdS boundary to the wormhole throat, which acts as a placeholder for the location of a microstate cap. The imaginary part, which determines the decay of the asymptotic wave function, only arises through the precise (small) coupling of AdS to asymptotically flat spacetime, congruent with earlier discussions \cite{Giusto:2004id,Giusto:2004ip,Giusto:2004kj,Jejjala:2005yu,Cardoso:2005gj,Bena:2020yii}. This is not surprising, as  the modes in the decoupling limit become normal modes without decay since $\text{AdS}_3$ acts as a confining box. The decay timescale in flat space then depends on the travel time down the throat that connects the decoupling region to flat space:	
\begin{equation}
 t _{\rm travel} \sim ({\rm Re}\, \omega)^{-1}\,, \qquad  t_{\rm decay} \sim ({\rm Im} \,\omega)^{-1} \sim  t _{\rm travel}^{-2\ell -3},
\end{equation}
with $\ell$ the $S^3$ wave number. This is in agreement with earlier observations and generic arguments for other compact objects based on crude estimates for the tunneling out of a potential barrier \cite{Cardoso:2019rvt}, as well as recent findings for microstate geometries which relate the travel time to the redshift or mass gap (the energy of the lowest excitation down the throat), $t_{\rm decay} \sim E_{\rm gap}^{-1}$ \cite{Bena:2020yii}.

\paragraph{Asymptotic AdS spacetimes and holography can give a window into flat space gravitational wave physics.} Within our setup, we relate the language of the CFT description dual to the $\text{AdS}_3$ (or BTZ) regions to recent results on black hole mimickers and exotic compact objects in flat space.  In section \ref{sec:flatAdSdictionary}, we show these descriptions are equivalent and we give the explicit dictionary relating them:
\begin{itemize}
\item If we assume that the microstates of a black hole are similar to the black hole near the asymptotic $\text{AdS}_3$ boundary --- and hence approximate the BTZ black hole spacetime --- then it is possible to model the complicated ECO potential by the black hole potential with different boundary conditions in the deep IR (replacing the ingoing boundary condition at the horizon). In this way, both the asymptotic behaviour of the wave equation at late times (including e.g. the echo structure and its frequency modulation) and the (approximate) QNMs are determined by the effective ECO reflectivity coefficient $R$ \cite{Mark:2017dnq}. We will discuss precisely how this reflection coefficient is related to the holographic correlator and the details of the microstate geometry.
This shows how the phenomenological description of GW echoes directly probes the field theory dual for the controllable setup of supersymmetric three-charge  black hole.

\item The structure of echoes in compact objects is sometimes catalogued by the distance of those corrections to the corresponding black hole's  horizon. Using our toy model and the (known) travel time in superstrata \cite{Bena:2019azk}, we find support for the idea that using the viewpoint that microstate geometries are `a black hole plus corrections', \emph{the microstructure would sit exponentially close to the horizon} --- in terms of the proper distance from the horizon (see section \ref{sec:qucorr}):
\begin{equation}
\Delta s \sim \exp (-\alpha S_{BH}),
\end{equation}
with $\alpha$ a number that depends on how far away from extremality the black hole is, and $S_{BH}$ the Bekenstein-Hawking entropy of the five-dimensional black hole (see eq.\ \eqref{eq:loglambdaS} below). Although it might seem impossible to have a classical geometry with such sub-planckian corrections in gravity, the microstate geometry itself is completely smooth and suffers no pathologies. Hence, microstate geometries offer a (semi-)classical description of seemingly quantum-sized corrections.\footnote{Microstate geometries can have sub-planckian modes, from waves being trapped near the evanescent ergosurface. However, such modes are beyond the supergravity approximation.} 
\end{itemize}

\paragraph{A  typical BH microstate will likely not lead to a discernible GW echo signal.} There have been many papers analyzing possible expected gravitational wave signals from echoes in black hole mimickers or microstates  (see for example \cite{Barack:2018yly,Cardoso:2017cqb,Cardoso:2017njb}). However, statistical arguments suggest that a \emph{typical} black hole microstate will in general be a quantum superposition of many such microstates and thus have no discernible echo structure --- see section \ref{sec:lambdaensemble}.

To address this question, we use our setup to perform an ensemble averaging over the parameter $\lambda$ and comment on the possibility of seeing echoes. 
The known instabilities of atypical microstate geometries \cite{Eperon:2016cdd} can be interpreted as leading to a final state that explores the phase space of all solutions \cite{Marolf:2016nwu} and generically ends up in a typical state. Transitions between states are reached through quantum tunneling \cite{Mathur:2008kg,Kraus:2015zda,  Bena:2015dpt}. Even though one could imagine that such transitions temporarily lead to a branch of the wave function describing the black hole that localizes on a microstate geometry, for instance by a GW burst following the collision and merger of two black holes \cite{Hertog:2017vod}, there is no reason to expect such a state to be long-lived and produce standard GW echoes. Rather, following relaxation, the response should be determined by the black hole behaviour, with the variance around the black hole mean of classical observables is exponentially suppressed. We show how after averaging over the wormhole parameter $\lambda$ in a microcanonical ensemble setup, the echo details are washed out, as the amplitude drastically decreases and can be exponentially suppressed; we also comment on the connection to recent discussions in four-dimensional gravity literature.
	
\subsection{Paper Structure}\label{sec:structure}

The rest of the paper is roughly divided into two parts.

In the first part, we focus on the  string theoretic calculation of the approximate propagator for the extremal BTZ wormhole and discussing its coupling to flat space. In section \ref{sec:BTZandWHs}, we review the BTZ black hole and introduce its corresponding Solodukhin-type wormholes as asymptotically $AdS_3$ spacetimes. 
Then, in section \ref{sec:AdS3WKB}, we use the hybrid WKB method of \cite{Bena:2019azk} to calculate the correlator and quasinormal modes of the extremal rotating wormhole. We show how this correlator in asymptotically $\text{AdS}_3$ spacetime actually captures flat space physics by explicitly showing their link using a matched asymptotic expansion in section \ref{sec:connectflat}.

In the second part of this paper, we discuss the link between AdS and flat space gravitational wave physics, as well as the feasability of observing echoes in gravitational wave signals. Section \ref{sec:flatAdSdictionary} reviews the language of flat space gravitational wave (echoes) and makes the link with AdS correlator physics explicit. Finally, in section \ref{sec:discuss} we discuss the location of the microstructure in relation to the black hole horizon, and our averaging procedure that indicates the echo signal is heavily damped in a typical microstate background.

\section{BTZ Black Holes and Wormholes}\label{sec:BTZandWHs}
Here, we introduce the asymptotically $\text{AdS}_3$ geometries that we will use extensively in the rest of the paper. First, we review the BTZ black hole. Then, we consider asymptotically BTZ wormholes which are taken and generalized from  \cite{solodukhin_restoring_2005}; we also extend this wormhole to the extremally rotating limit which will allow us to connect to the five-dimensional microstates discussed later in this paper. Along the way, we discuss a few basic properties of the wormholes that we will need later on.

\subsection{The BTZ Black Hole}
\begin{figure}[ht]
\begin{center}
   \includegraphics[]{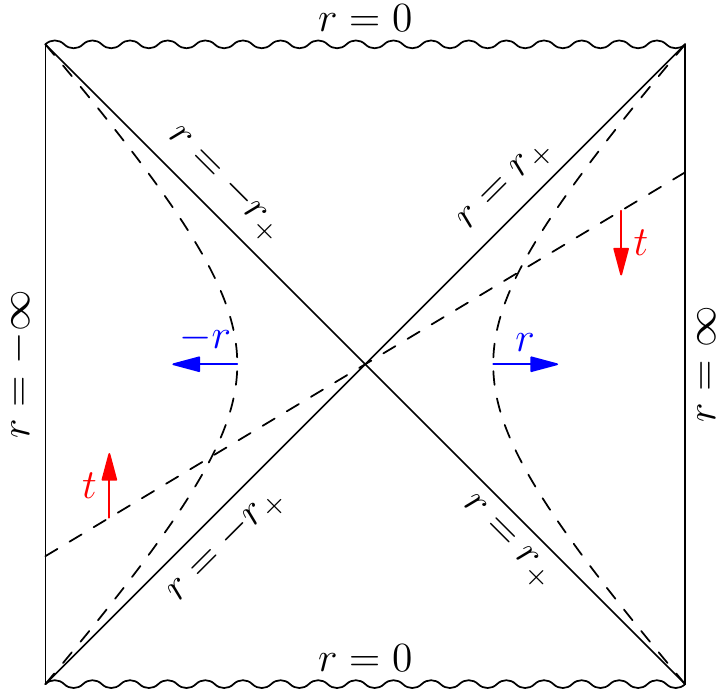}
\end{center}
\caption{Conformal diagram of the extended non-rotating BTZ black hole. The two asymptotically $\text{AdS}_3$ regions are near the lines $r=\pm \infty$. The singularity is at $r=0$. The dashed lines are lines of constant time and radius. }
\label{fig: black hole conformal diagrams}
\end{figure}

Our starting point is the BTZ black hole \cite{banados_black_1992,banados_geometry_1993} with metric:
\begin{align}
\label{eq:BTZmetric}
\begin{aligned}
\dd{s}^2 &= -X(r) \dd{t}^2 + \frac{ \dd{r}^2 }{Y(r)}+ r^2 \qty(\dd{\varphi} +N_{\varphi}(r)\dd{t})^2 \\
X(r)&=Y(r)= \frac{(r^2 - r^2_+) (r^2 - r^2_-)}{R^2 r^2}, \quad N_\varphi(r) = - \frac{r_+ r_-}{R r^2},
\end{aligned} \qquad \begin{cases}\begin{aligned} r &\in [r_+,\infty) \\
t &\in (-\infty,\infty) \\
\varphi &\sim \varphi + 2 \pi,
\end{aligned}
\end{cases}
\end{align}
where $R$ is the AdS radius. These coordinates cover the patch outside the horizon ($r>r_+$). One can use Kruskal coordinates in the standard way to maximally extend the spacetime, obtaining two outside (left and right) regions which each have a separate asymptotic $\text{AdS}_3$ region --- this can be interpreted as two entangled black holes connected by a non-traversible Einstein-Rosen bridge \cite{Maldacena:2013xja}; see also figure \ref{fig: black hole conformal diagrams}.
The asymptotic mass and angular momentum of the black hole, in units where $8G_3=1$, are
\begin{align}
M = \frac{r^2_+ + r^2_-}{R^2}, \qquad J = \frac{2r_+ r_-}{R},
\end{align}
while the temperature and the entropy of the black hole are given by
\begin{align}
T=\frac{r^2_+ - r^2_-}{2 \pi R^2 r_+}, \qquad S=4\pi r_+ .
\label{eq: temp and entropy of BTZ}
\end{align}
In the holographically dual $\text{CFT}_2$, the black hole corresponds to a thermal state with the same temperature. There are two special cases of interest we highlight for further use in this paper:
\begin{itemize}
\item $r_- = 0$  ($J=0$): the non-rotating, static BTZ black hole (see figure \ref{fig: black hole conformal diagrams}).
\item $r_+ = r_- $ ($J=MR$): the extremal BTZ black hole.
\end{itemize}

\subsection{Wormholes with BTZ Asymptotics}

By changing the metric factor $X(r)$ in (\ref{eq:BTZmetric}), we can alter the BTZ geometry to be a Solodukhin wormhole. For the extremal case, we will also need to change $Y(r)$. We will discuss the two cases described earlier: the non-extremal non-rotating wormhole $(r_-=0)$ and extremal rotating wormhole ($r_+ = r_-$).


\subsubsection{The non-rotating Solodukhin wormhole}\label{sec:DSWHnonrot}

It is straightforward to modify the non-rotating BTZ metric (\ref{eq:BTZmetric}) with $r_-=0$ into a wormhole as proposed originally by Solodukhin \cite{solodukhin_restoring_2005}:
\begin{align}
\begin{aligned}
\label{eq:WH3metric} \dd{s}^2 &= -X(r) \dd{t}^2 + \frac{\dd{r}^2}{Y(r)} + r^2 \dd{\varphi}^2 \\
X(r)&=\frac{r^2 -r_\lambda^2}{R^2},\quad    Y(r) = \frac{r^2 -r_+^2}{R^2},\\
r_\lambda^2 &\equiv r_+^2(1-\lambda^2),
\end{aligned} \quad \begin{cases}\begin{aligned} r &\in (-\infty,-r_+]\cup [r_+,\infty) \\
t &\in (-\infty,\infty) \\
\varphi &\sim \varphi + 2 \pi,
\end{aligned}
\end{cases}
\end{align}
This is a one-parameter family of modifications parametrized by the small dimensionless parameter $\lambda>0$. Just as in the original wormhole of \cite{solodukhin_restoring_2005}, when $r\gg r_+$ the metric looks almost like that of the original black hole since $\lambda$ is small. However, the global structure of spacetime is very different: there is no longer a horizon at $r=r_+$ since $X(r_+)>0$. Rather, at this position, we smoothly glue a second copy of this spacetime (with $r<-r_+$), thus creating a traversible wormhole. See section \ref{sec:scalarwave} for more details on how the two copies are ``glued'' together in practice.
We discuss the physical significance of the parameter $\lambda$ and its relation to quantum {corrections to} the corresponding BTZ black hole in section \ref{sec:qucorr}.

The stress energy tensor that supports this geometry is given by:
\begin{align}
\begin{aligned}
T_{\mu \nu} = R_{\mu\nu} - \frac12 R g_{\mu\nu} + \Lambda g_{\mu\nu} = \lambda^2
\mqty(
0 & 0 & 0 \\ 
0 & -\displaystyle{ r_+^2 \over \rho^2(\rho^2 + \lambda^2r_+^2)} & 0\\ 
0 & 0 & \displaystyle{ (1-\lambda^2)r^4_+(\rho^2 + r_+^2) \over R^2(\rho^2 + \lambda^2r_+^2)^2}),
\end{aligned}
\label{eq: stress energy tensor wormhole}
\end{align}
where we have defined $\rho^2 \equiv r^2 - r^2_+$.
As can be expected, the wormhole is no longer a vacuum $\text{AdS}_3$ solution to Einstein gravity (in contrast to the BTZ metric), but rather requires the presence of exotic matter with negative pressure. We can also calculate the stress tensor in the holographically dual boundary $\text{CFT}_2$, following the standard prescription \cite{Balasubramanian:1999re,Skenderis:2002wp}:
\begin{align}
2 \pi R T^\text{holo}_{\mu \nu} =  \pi{ r_+^2 \over R^2}\mqty(1 & 0 \\ 0 & R^2(1-2\lambda^2 )).
\end{align}
From this we see that the holographic stress tensor has $O(\lambda^2)$ corrections compared to the corresponding (BTZ) black hole.

The parameter $\lambda$ determines the length of the wormhole. To see this, we can define the tortoise coordinate $r_*(r)$ for $r>r_+$ as:
\begin{align}
\label{eq:WHnonrottortoise} r_*^{(r>r_+)}(r) &= \int^r_{r_+} \dd{r'} {1 \over \sqrt{X(r')Y(r')}} \nonumber \\
&= {R^2 \over r_+ \sqrt{1-\lambda^2}} \qty[K\qty(1 \over 1-\lambda^2) -F\qty(\arcsin\qty(r \over r_+);{1 \over 1-\lambda^2})],
\end{align}
where $K(\cdot)$ is a complete elliptic integral of the first kind and $F(\cdot;\cdot)$ is an elliptic integral of the first kind. The tortoise coordinate can then be extended over the entire wormhole spacetime as:
\begin{align}
\label{eq:WHnonrottortoisecomplete} r_*(r) = \theta(r-r_+) r_*^{(r>r_+)}(r) - \theta(-r-r_+) r_*^{(r>r_+)}(-r),
\end{align}
where $\theta(\cdot)$ is a Heaviside function. The tortoise coordinate ranges continuously in $[-L_\lambda/2,L_\lambda/2]$, where the asymptotic $\text{AdS}_3$ boundary $r\rightarrow+\infty$ is at $L_\lambda/2$, with:
\begin{align}
\label{eq:r*bnonrot} L_\lambda = 2{R^2 \over r_+ \sqrt{1-\lambda^2}} \qty[K\qty(1 \over 1-\lambda^2) + i K\qty(-{\lambda^2\over 1-\lambda^2})] =  \frac{R^2}{r_+} \log \frac{16}{\lambda^2}+O(\lambda^2)\,.
\end{align}
We call $L_\lambda$ the wormhole throat length. It corresponds to the travel time from one AdS boundary to the boundary on the other side, as measured by an observer at the asymptotic boundary of AdS (either one). For that reason, it was dubbed the `optical length' in \cite{solodukhin_restoring_2005}.

\begin{figure}[!ht]
\begin{center}
\begin{subfigure}[c]{0.48\textwidth}
	\centering
\includegraphics{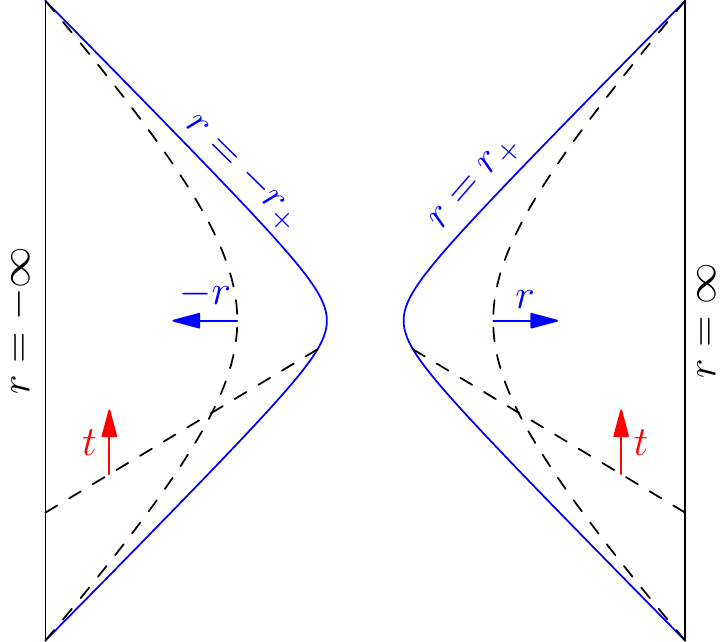}
\end{subfigure}
\hspace{.01\textwidth}
\begin{subfigure}[c]{0.48\textwidth}
	\centering
	\includegraphics{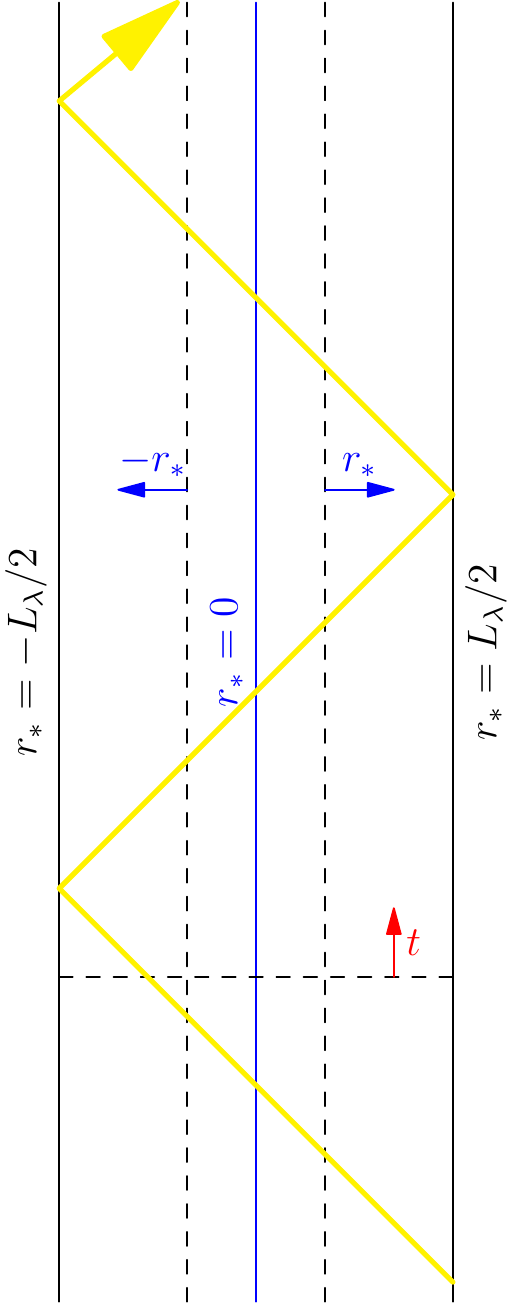}
\end{subfigure}
\end{center}
\caption{Conformal diagrams of the Lorentzian non-rotating wormhole in terms of $r$ (left) and $r_*$ (right) coordinates. In the $r$ coordinate we see that the wormhole is created by a left and a right BTZ wedge glued together (the blue solid line) slightly outside of the BTZ horizon. The $r_*$ coordinate, which ranges continuously in $[-L_\lambda/2,L_\lambda/2]$, makes the causal structure more apparent: the glued surfaces $r=\pm r_+$ (on the left) both coincide with the surface $r_*=0$ (on the right). The gluing is smooth and a radial light-ray (yellow) can explore the entire wormhole by bouncing infinitely many times between the two boundaries.}
\label{fig: non-rot wh}  
\end{figure}

The finiteness of the range of the tortoise coordinate distinguishes the wormhole from the black hole (for which the corresponding tortoise coordinate goes to $r_*\rightarrow -\infty$ at the horizon). In figure \ref{fig: non-rot wh} we give conformal diagrams of the non-rotating wormhole using, resp., $r$ and $r_*$ as radial coordinates. 

The QNMs of this non-rotating wormhole were determined by a WKB method in \cite{solodukhin_restoring_2005}; we will focus on the extremal rotating limit from now on.

\subsubsection{The extremal Solodukhin wormhole}\label{sec:extDSWH}

As we are interested in modelling extremally rotating BTZ black holes in string theory, we can also construct a wormhole based on the extremal BTZ black hole with $r_-=r_+$:
\begin{align}
\begin{aligned}
\label{eq:WH3extmetric} \dd{s}^2 &= -X(r) \dd{t}^2 + \frac{\dd{r}^2}{Y(r)} + r^2 \left(\dd{\varphi} -\frac{r_+^2}{Rr^2}\dd{t}\right)^2 \\
X(r)&={(r^2 -r_\lambda^2)^2 \over R^2r^2},\quad    Y(r) = {(r^2-r_\lambda^2)(r^2 -r_+^2) \over R^2r^2},\\
r_\lambda^2 &\equiv r_+^2(1-\lambda^2),
\end{aligned} \quad \begin{cases}\begin{aligned} r &\in (-\infty,-r_+]\cup [r_+,\infty) \\
t &\in (-\infty,\infty) \\
\varphi &\sim \varphi + 2 \pi,
\end{aligned}
\end{cases}
\end{align}
We chose to construct the extremal wormhole in such a way that the gluing point $r=r_+$ is a \emph{simple} zero of $Y(r)$.\footnote{This is a choice; other constructions are also possible. We made this choice since when $r=r_+$ is a simple zero of $Y(r)$, it is somewhat easier to integrate the tortoise coordinate expression.} The discussion of the extremal wormhole proceeds very similarly to the above discussion for the non-rotating wormhole; in particular, the wormhole throat at $r=r_+$ is glued to a second copy of the spacetime with $r<-r_+$.

The wormhole throat length is qualitatively different in the extremal limit. To see this, once again define the tortoise coordinate:
\begin{align}\label{eq:WH3exttortoise}
r_*^{(r>r_+)}(r) & = \int_{r_+}^r dr' \frac{1}{\sqrt{X(r')Y(r')}}\\
\nonumber &= \left(\frac{R^2}{r_+}\frac{1}{\lambda^2} + \mathcal{O}(\lambda^0,\log\lambda)\right) +\frac{R^2}{r_+^2-r_\lambda^2}\left( r \sqrt{\frac{r^2-r_+^2}{r^2-r_\lambda^2}} - r_+ E\left( \sin^{-1}\frac{r}{r_\lambda}, \frac{r_\lambda^2}{r_+^2}\right) \right),
\end{align}
where $E(\cdot,\cdot)$ is the elliptic integral of the second kind.
The asymptotic $\text{AdS}_3$ boundary is located at $r_*=L_\lambda/2$ with:
\be \label{eq:WH3extdefthroatlength}  L_\lambda = 2\frac{R^2}{r_+}\frac{1}{\lambda^2} + \mathcal{O}(\lambda^0,\log\lambda).\ee
Since the time coordinate $t$ corresponds to the asymptotic time of the D1-D5-P black hole metric (see section\ \ref{sec:connectflat}), the throat length $L_\lambda$ is now a measure of the travel time from the AdS boundary to other boundary, both as measured by an observer at the AdS boundary, and an observer in the asymptotically flat spacetime of the D1-D5-P black hole.

The throat length is polynomially dependent on (the inverse of) $\lambda$, as opposed to the logarithmic dependence for the non-rotating wormhole (\ref{eq:r*bnonrot}). This behavior is typical for redshifts associated to extremal black hole horizons.
When $\lambda^2\ll (r-r_+)/r_+\ll 1$, we can write the approximate inverse relation:
\be \label{eq:WH3extrxapprox} r(r_*) \approx \frac{R^2}{4}\frac{1}{L_\lambda/2-r_*}\left(1+\sqrt{1+16\frac{r_+^2 (r_*-L_\lambda/2)^2}{R^4}}\right) ,\ee
From here we see that for $r-r_+\sim \mathcal{O}(\lambda^0)$, we must also have $r_*\sim \mathcal{O}(L_\lambda)\sim \mathcal{O}(\lambda^{-2})$. From (\ref{eq:WH3exttortoise}), we can also see that $r_*\sim \mathcal{O}(\lambda^0)$ when $r-r_+\sim \mathcal{O}(\lambda^6)$.

\subsection{The Scalar Wave Equation}\label{sec:scalarwave}
We will be considering a probe scalar $\Phi$ on the BTZ metric background (\ref{eq:BTZmetric}) or the wormhole metrics (\ref{eq:WH3metric}) and (\ref{eq:WH3extmetric}), with general equation of motion:
\be \nabla^2 \Phi = m^2 \Phi.\ee
We can solve this by a separation of variables:
\be \label{eq:scalar3Dsep} \Phi(t,r,\varphi) = e^{-i\omega t + i k \varphi} \Phi_r(r),\ee
where the radial function $\Phi_r(r)$ must satisfy the differential equation:
\be \label{eq:scalar3Dradial} \Phi_r'' + \left( \frac{1}{r} + \frac{X'}{2X} + \frac{Y'}{2Y}\right) \Phi_r' + \left( \frac{\omega^2}{X Y} - \frac{k^2}{r^2 Y}-\frac{m^2}{Y}+k N_\varphi\frac{2\omega+k N_\varphi}{X Y}\right) \Phi_r = 0.\ee


\paragraph{Relating wormhole to BTZ solutions and matching conditions}
We will be interested in finding the scalar wave solutions to (\ref{eq:scalar3Dradial}) in both the black hole and wormhole metrics. For the BTZ black hole, the solutions to (\ref{eq:scalar3Dradial}) are explicitly known. Here, we will discuss how the solutions to the corresponding wormhole are related to these black hole solutions.

Let us work in the non-rotating asymptotically $\text{AdS}_3$ wormhole given by (\ref{eq:WH3metric}) for definitiveness. We work with the wormhole tortoise coordinate $r_*^{\text{WH}}(r)$ introduced in (\ref{eq:WHnonrottortoise}), {which} can be expanded over the entire wormhole as in (\ref{eq:WHnonrottortoisecomplete}).
Note that $r_*^{\text{WH}}=0$ indicates the center of the wormhole throat or the ``gluing point''. The corresponding (non-rotating) black hole (\ref{eq:BTZmetric}) has tortoise coordinate:
\be \label{eq:xBHnonrot} r_*^{\text{BH}}(r) = \frac{R^2}{2r_+}\log \left(\frac{r-r_+}{r+r_+}\right) + \frac{L_\lambda}{2},\ee
where we have used the constant of integration to set $r_*^{\text{BH}}(r\rightarrow \infty) = r_*^{\text{WH}}(r\rightarrow\infty) = L_\lambda/2 + \mathcal{O}(\lambda^2)$, with the throat length defined in (\ref{eq:r*bnonrot}).

Now, the radial scalar wave equation (\ref{eq:scalar3Dradial}) can be rewritten using $\Phi_r(r) = h(r) \phi(r_*)$, with $h(r) = \sqrt{R/r}$, as:
\be \label{eq:radialeqBHandWH} \partial_{r_*}^2 \phi(r_*) - V(r_*) \phi(r_*) = 0,\ee
for both the black hole and wormhole geometries (using the respective tortoise coordinate for each). Now, we can easily see that on one side of the wormhole, $V^{\text{WH}}(r_*^{\text{WH}}(r)) = V^{\text{BH}}(r_*^{\text{BH}}(r)) + \mathcal{O}(\lambda^2)$, so that over the entire wormhole:
\be \label{eq:fullWHpotential} V^{\text{WH}}(r_*^{\text{WH}}) = \theta(r_*^{\text{WH}}) V^{\text{BH}}\left(r_*^{\text{BH}}=r_*^{\text{WH}}\right) + (r_*^{\text{WH}}\rightarrow -r_*^{\text{WH}}) + \mathcal{O}(\lambda^2).\ee
Thus, the general solution $\phi^{\text{BH}}(r)$ to the black hole radial equation (\ref{eq:radialeqBHandWH}) are also the general solutions (to $\mathcal{O}(\lambda^2)$) to the wormhole radial equation, so $\phi^{\text{WH}}(r) = \phi^{\text{BH}}(r)$ on the right side of the wormhole, and $\phi^{\text{WH}}(-r) = \phi^{\text{BH}}(r)$ on the other side. Additionally, the general solution of the wormhole radial equation (and its first derivative) must be continuous at the wormhole throat radius $r_t$. From (\ref{eq:fullWHpotential}), we see that the wormhole throat $(r_*^{\text{WH}})_t=0$ is at $(r_*^{\text{BH}})_t = 0$, which translates in the coordinate $r$ as (keeping the leading order difference with $r_+$):
\be \label{eq:rstarnonrotWH} r_t^{\text{(non-rot})} = r_+\left(1 + \frac{\lambda^2}{8}\right) + \mathcal{O}(\lambda^4).\ee
The full solution to the scalar wave equation thus must satisfy:
\be \label{eq:rtmatchingcond} \phi^{\text{WH}}(r_t) = \phi^{\text{WH}}(-r_t), \qquad \partial_r\phi^{\text{WH}}(r_t) = \partial_r\phi^{\text{WH}}(-r_t).\ee

For the extremal wormhole (\ref{eq:WH3extmetric}) and its corresponding extremal black hole (\ref{eq:BTZmetric}), a similar calculation can be done. This is a bit trickier as the extremal black hole tortoise coordinate cannot be exactly inverted to give $r(r_*)$. However, since only an approximate relation to $\mathcal{O}(\lambda^2)$ is needed, we can still find:
\be \label{eq:rstarextWH} r_t^{\text{(ext)}}  = r_+\left(1 + \frac{\lambda^2}{4}\right).\ee

\paragraph{Tortoise and other global wormhole coordinates}
The tortoise coordinate $r_*$ has a physical interpretation as measuring the travel time of a light ray. Thus, its finite range (see (\ref{eq:r*bnonrot}) and (\ref{eq:WH3extdefthroatlength})) for the wormhole indicates {that} the entire wormhole geometry is accessible by null geodesics in a finite (coordinate) time $t$. For example, a light ray can travel radially from the right boundary, through the wormhole, reflect on the left boundary and return to the original point in a time $2L_\lambda$. See also figure \ref{fig: non-rot wh}. Thus, we can view the wormhole as a rudimentary toy model of a smooth, horizonless black hole microstate; the left side of the wormhole, including the asymptotic AdS boundary (where we assume reflecting boundary conditions), acts as a ``cap'' which reflects any infalling matter back out in finite (but long) time, controlled by the wormhole length $L_\lambda$.

The tortoise coordinate $r_*$ is not the unique coordinate for which the wormhole potential takes the symmetric form (\ref{eq:fullWHpotential}). In section \ref{sec:AdS3WKB}, to facilitate the hybrid WKB analysis, we will introduce a coordinate $x$, which is also globally defined over the entire wormhole and in which the radial equation also takes the Schr\"odinger form (\ref{eq:radialeqBHandWH}) with a symmetric potential such as (\ref{eq:fullWHpotential}). The coordinate $x$ moreover will be chosen to have an infinite range and satisfy $\lim_{x\rightarrow \infty}V(x) = 1$; this requires a different choice for the function $h(r)$ than what we choose above.

\section{The Hybrid WKB Approximation}\label{sec:AdS3WKB}
We discussed the equation of motion for a probe scalar field $\Phi$ in the asymptotically $\text{AdS}_3$ spacetimes of the BTZ black hole and wormholes in section \ref{sec:scalarwave}. Any such bulk $\text{AdS}_3$ scalar field will be holographically dual to a scalar operator $\mathcal{O}$ on the two-dimensional CFT that lives on the ``boundary'' (at spatial infinity) of $\text{AdS}_3$ \cite{Aharony:1999ti,skenderis_lecture_2002}. In this section, we will calculate the two-point correlation function of this holographic scalar operator $\mathcal{O}$.

Below, we will introduce a radial coordinate $x$ for which the $\text{AdS}_3$ boundary is at $x\rightarrow \infty$.
Solutions of the bulk $\text{AdS}_3$ scalar field $\Phi$ behave near the boundary in these coordinates as (we suppress angular $\varphi$ dependence for simplicity):
\be \Phi(t,x) \approx e^{-it\omega}\left(\beta(\omega) e^{\mu x} + \alpha(\omega) e^{-\mu x}\right).\label{eq:alphabeta_first}\ee
This asymptotic expansion of $\Phi$ can be used to calculate correlation functions of the holographically dual scalar operator $\mathcal{O}$ in the boundary CFT. In particular, we will be interested in the retarded two-point correlator $G^R(t) \equiv \theta(t) \langle [\mathcal{O}(t),\mathcal{O}(0)]\rangle$, which for a linearized analysis of the scalar field is related to the ratio of coefficients of the asymptotic expansion of the bulk field $\Phi$ as \cite{skenderis_lecture_2002,Bena:2019azk}:
\be \label{eq:RomegadefWKB} \mathcal{R}(\omega) \equiv -i \int dt e^{-i t\omega} G^R(t) =  \frac{\alpha (\omega)}{\beta (\omega)}.\ee
In order to calculate the holographic correlator (\ref{eq:RomegadefWKB}) in the extremal wormhole toy model, we will use the hybrid WKB technique developed by \cite{Bena:2019azk}. First, we will give a brief overview of the method in section \ref{sec:AdS3WKB:method} and the main steps in applying it to the systems at hand. Then we will review the calculation of the free scalar correlator in the extremal BTZ black hole background in section \ref{sec:WKBBTZBH}, and then redo the calculation for the the extremal BTZ wormhole in section \ref{sec:AdS3WKB:ext}. Finally, we give some brief remarks on the expectations for the free scalar position-space propagator $G^R(t)$ in section \ref{sec:WHPSprop}.

\subsection{Method Outline}\label{sec:AdS3WKB:method}
The hybrid WKB approximation works in a number of steps. First, we apply the traditional WKB method to find an approximate solution to the wave equation.
In order to proceed with finding the WKB approximation to the propagator, one needs to analyze the asymptotic structure of the approximate solution around the boundary.
However, this is generally a hard task, because the WKB solution is not written in a basis that is conducive to asymptotic analysis.
This is where the second step of the hybrid WKB technique is employed.
One finds a new potential that asymptotically approaches the original potential and for which the wave equation can be solved exactly.
Then the WKB solution is matched with the asymptotic solution, which allows one to efficiently analyze the asymptotic structure of the approximate solutions to the wave equation and ultimately extract the propagator.

Practically speaking, we will divide the method into three parts, which we will now discuss.

\subsubsection{Setting up the radial equation and potential}
We wish to split the radial part of the scalar as:
\begin{equation}
\Phi_r(r) = h(r) \phi(r) \,,
\end{equation}
and we perform a coordinate transformation $ r \rightarrow x(r)$ to write the radial wave equation (\ref{eq:scalar3Dradial}) for $\phi$ in the Schr{\"o}dinger form:
\begin{equation}\label{eq:WKBSchr}
\frac{\partial^2}{\partial x^2} \phi(x)- V(x) \phi(x)=0 \,.
\end{equation}
The absence of a term proportional to $\partial_x\phi$ leads to the condition:
\begin{equation}\label{eq:WH3feq}
h(r) = \frac{1}{\left(r^2X(r)Y(r)\left(\frac{\partial x(r)}{\partial r}\right)^2\right)^{1/4}} \,.
\end{equation}
We are still free to choose $x(r)$, which will affect the form of the potential $V(x)$.
To align with the analysis in \cite{Bena:2019azk}, we choose $x(r)$ such that the potential approaches a particular constant in the UV:
\begin{equation} \label{eq:WKBVasympt}
\lim_{x\rightarrow \infty} V(x) = 1+m^2 R^2 \equiv \mu^2 \,.
\end{equation}
where we have defined the dimensionless parameter $\mu$.
This can be achieved by fixing the function $x(r)$ such that:
\be \label{eq:WKBfr} h(r) = \frac{R}{r}~.\ee

The following condition needs to be met in order for the WKB approximation to be valid
\begin{equation}\label{eq:wkbapprox}
\left|\left(V(x)\right)^{-\frac{3}{2}}\frac{\partial V(x)}{\partial x}\right| \ll 1 \,.
\end{equation}
This quantity should be evaluated away from the turning points, defined by $V(x_\text{turn})=0$, where the left side diverges.

\subsubsection{Asymptotic solutions}
For large $x$, the potential asymptotes to a potential $V_\text{asymp}(x)$:
\be V(x) \sim V_\text{asymp}(x),\ee
that we will make precise below. The solutions to the wave equation (\ref{eq:WKBSchr}) with this potential $V_\text{asymp}(x)$ are given by $\phi_\text{asymp}(x)$ with the following structure
\begin{equation}\label{eq:solutionphi}
\phi_\text{asymp}(x)= c_1 \phi_\text{asymp}^\text{grow}(x)+c_2 \phi_\text{asymp}^\text{dec}(x)\,,
\end{equation}
where in the UV we have defined the two linearly independent solutions as those that behave at large $x$ as:\footnote{Note that the full radial function $\Phi_r$ in (\ref{eq:scalar3Dsep}) is still given by $\Phi_r(r) = \phi(x)R/r$.}
\begin{equation}\label{eq:WKBstructasymptphi}\begin{aligned}
\phi_\text{asymp}^\text{grow}(x) &= e^{\mu x}\left(1+\dots\right)\,,\\
\phi_\text{asymp}^\text{dec}(x) &=  e^{-\mu x}\left(1+\dots\right)\,,
\end{aligned}\end{equation}
where the omitted terms are subleading in $x$ at large $x$.
When $\mu$ is an integer, there are also pieces proportional to $x$ in the UV expansion; we will assume $\mu$ not to be an integer and analytically continue the final result to integer $\mu$.
Both $\phi_\text{asymp}^\text{grow}$ and $\phi_\text{asymp}^\text{dec}$ are real functions of $x$ (for real $r_+,k,\omega,\mu,R$).

\subsubsection{Calculation of the WKB propagator}
Now we have all ingredients to evaluate the propagator that follows from the hybrid WKB technique.
According to \cite{Bena:2019azk}, the propagator is given by
\begin{equation}
\mathcal{R} = \left(\mathcal{A}+\frac{\sqrt{3}}{2}\right)e^{-2I_+}-\frac{\phi_\text{asymp}^\text{grow}(x_+)}{\phi_\text{asymp}^\text{dec}(x_+)} \,.
\end{equation}
where
\begin{equation}
I_+ = -\mu x_+ + \int_{x_+}^\infty \left(\sqrt{|V(z)|}-\mu\right)dz \,,
\end{equation}
and the turning points $x_{\pm}$ are defined such that $V(x_\pm)=0$.
The expression for $\mathcal{A}$ depends on the number of turning points.
For a single turning point, we have
\begin{equation}
\mathcal{A} = \frac{i}{2} \textrm{sign}(\omega)  \,.
\end{equation}
For two turning points $x_-$ and $x_+$, we have
\begin{equation}\begin{aligned}
\mathcal{A} = \frac{1}{2} \tan(\Theta) \,, \qquad \Theta = \int_{x_-}^{x_+}\sqrt{|V(z)|}dz \,.
\end{aligned}\end{equation}
This will be the formula we will need to evaluate to find the wormhole propagator.
The general expression for $\mathcal{A}$ for more than two turning points can be found in an appendix of \cite{Bena:2019azk}.

\subsection{The BTZ Black Hole}\label{sec:WKBBTZBH}

It is useful to first review the hybrid WKB calculation applied to the extremal BTZ black hole itself. 

\subsubsection{Setting up the radial equation and potential}

We choose the radial coordinate:
\be x^\text{BH}(r) = \frac12 \log \frac{r^2-r_+^2}{r_+^2} - \frac12\frac{r_+^2}{r^2-r_+^2},\ee
such that the black hole potential is
\be\label{eq:BHpotential}
V^{\rm BH} (x^{\rm BH} (r_*)) =  \frac{(r^2-r_+^2)^2}{r^8}(2r^2r_+^2-3r_+^4+r^4\mu^2)+ \frac{R^2}{r^2}(k^2-R^2\omega^2)-\frac{R^2r_+^2}{r^4}2k(k-R\omega),
\ee
for which indeed $V(x\rightarrow\infty)=\mu^2$.
It is plotted along with the wormhole potential in figure \ref{fig:extWHpotentialx}.

\subsubsection{Asymptotic solutions}

The explicit analytic solutions $\phi_\text{asymp}^\text{dec}(x,)\phi_\text{asymp}^\text{grow}(x)$ for the extremal BTZ black hole are known in terms of Whittaker $M,W$ functions (see (\ref{eq:solAdSregion})), however, we do not actually need their expressions. Remarkably, as we will show below, we can use the WKB method to find an approximation for the wormhole propagator in terms of the black hole one without using the form of the explicit solutions $\phi_\text{asymp}$.

\subsubsection{Calculation of the WKB propagator}

The BTZ potential has only one turning point, so one finds
\begin{equation} \label{eq:WKBBTZprop}
\mathcal{R}^\text{BH} \approx \left(\frac{i}{2}\textrm{sign}(\omega)+\frac{\sqrt{3}}{2}\right)e^{-2I_+}-\frac{\phi_\text{asymp}^\text{grow}(x_+)}{\phi_\text{asymp}^\text{dec}(x_+)} \,.
\end{equation}
As noted earlier, the exact solutions $\phi_\text{asymp}^\text{grow,dec}$ are real and so is $I_+$.
This allows us to write
\begin{equation}\begin{aligned}
\textrm{Re}\left(\mathcal{R}^\text{BH}\right) &\approx \frac{\sqrt{3}}{2}e^{-2I_+}-\frac{\phi_\text{asymp}^\text{grow}(x_+)}{\phi_\text{asymp}^\text{dec}(x_+)} \,, \\
\textrm{Im}\left(\mathcal{R}^\text{BH}\right) &\approx \frac{1}{2}\textrm{sign}(\omega) e^{-2I_+}\,.
\end{aligned}\end{equation}
This will be useful when we calculate the WKB approximation for the wormhole propagator later on. Note that using the solutions $\phi_\text{asymp}$ one can compute the extremal BTZ propagator; this was done in \cite{Bena:2019azk}.


\subsection{The Extremal Wormhole}\label{sec:AdS3WKB:ext}
We can now apply the method above to find the approximate propagator in the Solodukhin extremal wormhole metric (\ref{eq:WH3extmetric}).

\subsubsection{Setting up the radial equation and potential}
We choose to work with the following coordinate:
\begin{equation}\label{eq:xcoordext}
x(r) = \frac{r_\lambda^2}{r_+^2-r_\lambda^2}\sqrt{\frac{r^2-r_+^2}{r^2-r_\lambda^2}} +    \log \left(\sqrt{\frac{r^2-r_\lambda^2}{r_\lambda^2}}+\sqrt{\frac{r^2-r_+^2}{r_\lambda^2}}\right)- \frac{1}{\sqrt{3}}\frac{1}{\lambda^2} \, ,
\end{equation}
for which $h(r)$ satisfies (\ref{eq:WKBfr}) and $x(r_t)=O(\log\lambda)$ (where $r_t$ is the matching radius given by (\ref{eq:rstarextWH})); the range of $x$ coordinate is $-\infty$ to $\infty$ and spans the whole wormhole.
The potential in this $x$ coordinate is:
\be\label{eq:Vapprox} V(x(r),\lambda) =V^{\rm BH}(x) + \mathcal{O}(\lambda^2),\ee
with $V^{\rm BH}(x)$ given in \eqref{eq:BHpotential} such that again $V(x\rightarrow\infty)=\mu^2$. At the wormhole throat $x=0$:
\be V(x=0) = -\frac{R^2}{r_+^2}(k-R\omega)^2 + \mathcal{O}(\lambda^2).\ee

To check that the condition (\ref{eq:wkbapprox}) is satisfied, we note that away from the turning point of the potential, $r-r_+\ll r_+$ (see also the discussion below, around (\ref{eq:derintegrandWHext})). Expanding (\ref{eq:wkbapprox}) to second order in $r-r_+$, we get:
\be \left|\left(V(x)\right)^{-\frac{3}{2}}\frac{\partial V(x)}{\partial x}\right| = \left| \frac{(R \omega - 3k)}{(R \omega - k)^2}\left( \frac{6(r-r_+)}{R}\lambda^2 + \frac{8(r-r_+)^2}{R r_+}\right) \right| + \mathcal{O}((r-r_+)^2\lambda^2,(r-r_+)^3).
\ee
The condition that this is much smaller than unity gives us the condition:
\be \label{eq:WKBomegacond} (R \omega - k)^2 \gg 12 |k|\lambda^2 \left(\frac{r-r_+}{R}\right) = 24|k|  \frac{R}{L_\lambda}\left(\frac{r-r_+}{r_+}\right) ,\ee
so we need $(R\omega  - k)$ to be sufficiently large so that the WKB approximation is valid.

\subsubsection{Asymptotic solutions}

\begin{figure}[ht]\centering
 \includegraphics[width=0.8\textwidth]{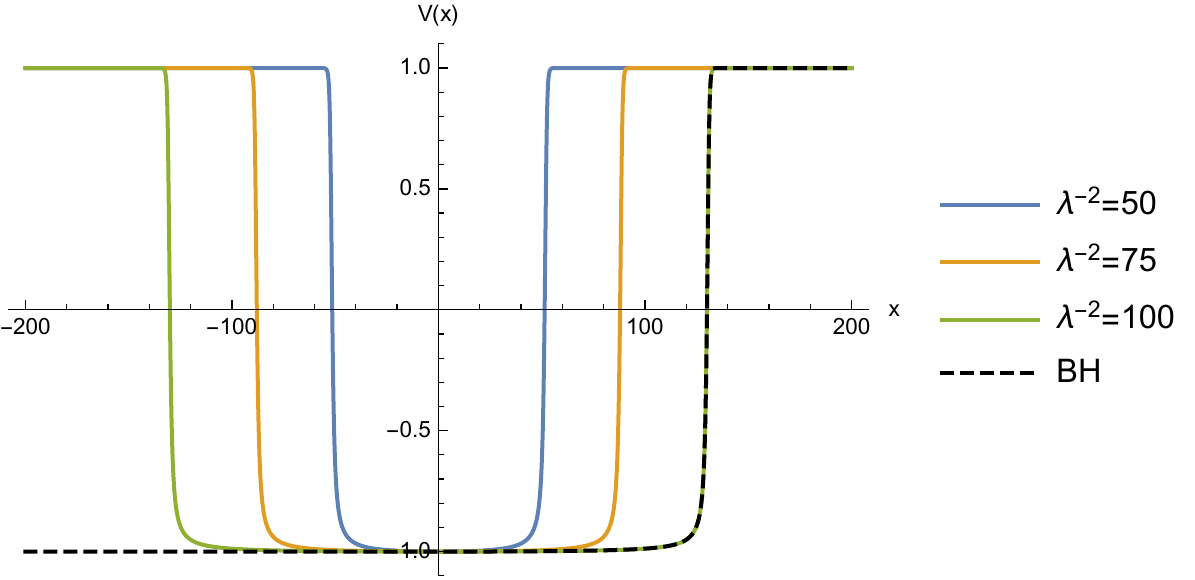}
 \caption{The wormhole potential $V(x)$, plotted for three different values of $\lambda$, and $R=r_+ = \omega=1$, $k=0$. The dotted line is the black hole potential $V_{\text{asymp}}(x-1/\lambda^2)$ for $\lambda^{-2}=100$, showing graphically the validity of the approximation (\ref{eq:Vdouble}).}
 \label{fig:extWHpotentialx}
\end{figure}

Because $x^\text{BH}(r_t) = -1/\lambda^2+\mathcal{O}(\log\lambda)$ (whereas for the wormhole, $x(r_t) = \mathcal{O}(\log\lambda)$), we can conclude that the wormhole potential is well approximated by two copies of the shifted asymptotic black hole potential:
\be \label{eq:Vdouble} V(x,\lambda) = \theta(x) V_{\text{asymp}}\left(x - \frac{1}{\lambda^2}\right) + (x\rightarrow -x) + \mathcal{O}(\log\lambda).\ee
This is, of course, the analogous relation in the $x$ coordinate for the extremal wormhole to (\ref{eq:fullWHpotential}). We have plotted an example of this potential in figure \ref{fig:extWHpotentialx}.

\subsubsection{Calculation of the WKB propagator}
The wormhole potential is essentially two copies of the black hole potential, leading to two turning points, so $\mathcal{A}$ will differ with respect to the BTZ propagator (\ref{eq:WKBBTZprop}).
However, due to \eqref{eq:Vapprox} we can see that $I_+^\text{wormhole}=I_+^\text{BH}+\mathcal{O}(\lambda^2)$, and thus we can write 
\begin{equation}\begin{aligned}\label{eq:RWKB}
\mathcal{R}^\text{WH} &\approx \textrm{Re}\left(\mathcal{R}^\text{BH}\right)+2 \,\textrm{sign}(\omega) \, \mathcal{A} \, \textrm{Im}\left(\mathcal{R}^\text{BH}\right) \,,
\end{aligned}\end{equation}
where $\approx$ now denotes both inaccuracy due to the WKB approximation and corrections of order $\lambda^2$. Thus, as anticipated, we do not need the explicit form of $\phi_\text{asymp}$ to be able to write down our wormhole propagator in function of the black hole one.

To get a rough approximation for $\Theta$, we can use the change of integration variables to the tortoise coordinate $r_*$ defined by (\ref{eq:WH3exttortoise}):
\begin{equation}\begin{aligned}\label{eq:thetaWH3ext}
\Theta &= \int_{x_-}^{x_+} \sqrt{\left|V(x)\right|} dx
= \int_{r_*(x_-)}^{r_*(x_+)}  \sqrt{\left|V(x(r(r_*)))\right|} \frac{dx}{dr}\frac{dr}{dr_*} dr_* \,.
\end{aligned}\end{equation} For $\lambda\ll1$, this integral is dominated by a flat potential in the wormhole throat. To see this, note that
\begin{equation}\label{eq:Vxatthroat}
\lim_{r\rightarrow r_+}  \sqrt{\left|V(x(r(r_*)))\right|} \frac{dx}{dr}\frac{dr}{dr_*} = \left|\omega-\frac{k}{R}\right| \,,
\end{equation}
and further that (for $x_+>x>0$, that is on the right side of the wormhole $r>r_+$):
\begin{equation}\label{eq:derintegrandWHext}
\frac{d}{d x} \left( \sqrt{\left|V(x(r(r_*)))\right|} \frac{dx}{dr}\frac{dr}{dr_*}\right) \approx \frac{8|k|}{R r_+^2}(r-r_+)^2\,.
\end{equation}
From the discussion of the extremal wormhole tortoise coordinate around (\ref{eq:WH3exttortoise}) and (\ref{eq:WH3extrxapprox}), we know that as long as $r_* \lesssim L_\lambda/2$, we have $r-r_+\ll r_+$ and so (\ref{eq:derintegrandWHext}) implies the potential remains very flat; thus, we can approximate (\ref{eq:thetaWH3ext}) very well by
\begin{equation} \label{eq:thetaextwh}
\Theta \approx \left|\omega-\frac{k}{R}\right| L_\lambda.
\end{equation}
Note that $R \omega-k$ needs to sufficiently large for the WKB approximation to be valid, see (\ref{eq:WKBomegacond}). The approximate value (\ref{eq:thetaextwh}) is indeed what we would expect intuitively from a potential that is very similar to that of a square well (see figure \ref{fig:extWHpotentialx}) with a depth of (\ref{eq:Vxatthroat}) and a length of $L_\lambda$.
The normal modes of the wormhole are the poles of $\tan\Theta$, which are then (approximately):
\be \label{eq:extBTZNMs} \omega_n = \frac{k}{R} + \left(n+\frac12\right)\frac{\pi}{L_\lambda}.\ee
The spacing between modes is to leading order:
\be \label{eq:extBTZNMspacing} \Delta \omega_n \equiv \omega_{n+1}-\omega_n = \frac{\pi}{L_\lambda},\ee
In this rough approximation, the propagator is then given by:
\begin{equation}\begin{aligned}\label{eq:extBTZfinalprop}
\mathcal{R}^\text{WH} &\approx \textrm{Re}\left(\mathcal{R}^\text{BH}\right)+\tan \left(\left|\omega-\frac{k}{R}\right| L_\lambda\right) \, \textrm{Im}\left(\mathcal{R}^\text{BH}\right) \,,
\end{aligned}\end{equation}
We restrict to the result for the QNMs to leading order in $\lambda$, as this is enough to for our purposes in the next sections. At next-to-leading order, the modes are no longer evenly spaced, as discussed for the non-rotating wormhole in \cite{solodukhin_restoring_2005}. 
We will briefly discuss the position-space propagator that follows from (\ref{eq:extBTZfinalprop}) below in section \ref{sec:WHPSprop}.

It is interesting to compare our (rough) result (\ref{eq:thetaextwh}) with the analysis of \cite{Bena:2019azk}, where the scalar correlator was calculated in the complicated $(1,0,n)$ superstratum metric (which asymptotes at the $\text{AdS}_3$ boundary as extremal BTZ). In particular, we can compare their result for $\Theta$ in the microstate cap regime  when the scalar mode frequency is high enough (see \cite{Bena:2019azk}, eq. (5.3)). The general structure they find for $\Theta$ and thus for the correlator matches our result qualitatively very well: the dominant piece in $\Theta$ is linear in $\omega$ and proportional to the microstate throat length $L_{micro}$ (with $L_{micro}\sim R_y b^2/a^2$, using the parameters in \cite{Bena:2019azk}). As they also find, this piece is responsible for echoes in the position space correlator at a time of order $\sim L_{micro}$ and integer multiplies of this (see \cite{Bena:2019azk}, figure 13). Thus, our very simple Solodukhin wormhole toy model captures this feature of the complicated superstratum geometry very well.

\subsection{Position-Space Propagator}\label{sec:WHPSprop}
We can use a few simple arguments to understand the general structure of the position-space propagator that follows from the frequency space result (\ref{eq:extBTZfinalprop}). Note that in \cite{Bena:2019azk}, the position-space propagator is discussed at length.

Consider the distributional identity:
\be \tan x = \sum_{n=1}^\infty 2(-1)^{n-1} \sin (2nx),\ee
and its Fourier transform:
\be \mathcal{F}(t) \equiv \frac{1}{2\pi}\int d\omega e^{-i\omega t} \tan \omega L_\lambda = i \sum_{n=1}^\infty (-1)^{n-1} \left(\delta(t+2 n L_\lambda) - \delta(t-2n L_\lambda)\right).\ee
Then, using the convolution theorem for Fourier transforms on the retarded propagator (\ref{eq:extBTZfinalprop}):
\begin{align}
 G^R_\text{WH}(\Delta t, \Delta \varphi) &\approx   G^0(\Delta t,\Delta \varphi) + (G^1*\mathcal{F})(\Delta t,\Delta \varphi),\\
 G^0(\Delta x) & \equiv -\frac{\theta(\Delta t)}{2\pi^2 } \sum_k e^{ik\Delta \varphi}\int d\omega e^{-i\omega \Delta t} \text{Re}(\mathcal{R}^\text{BH}), \\
  G^1(\Delta x) & \equiv -\frac{\theta(\Delta t)}{2\pi^2 } \sum_k e^{ik\Delta \varphi}\int d\omega e^{-i\omega \Delta t} \text{Im}(\mathcal{R}^\text{BH}),
 \end{align}
so that $G^R_\text{BH}(\Delta t,\Delta \varphi) = G^0(\Delta t,\Delta \varphi) + iG^1(\Delta t,\Delta \varphi)$, and the convolution gives:
\begin{align}\nonumber (G^1*\mathcal{F})(\Delta t,\Delta \varphi) &=    i \sum_{n=1}^\infty (-1)^{n-1} \int dt' G^1(\Delta t-t', \Delta \varphi) \left(\delta(t'+2 n L_\lambda) - \delta(t'-2n L_\lambda)\right)\\
&=  i \sum_{n=1}^\infty (-1)^{n-1} \left(G^1(\Delta t +2n L_\lambda) - G^1(\Delta t-2n L_\lambda)\right),
\end{align}
so that we have the relation:
\be \label{eq:naiveWHGR} G^R_\text{WH}(\Delta t, \Delta \varphi) \approx G^0(\Delta t,\Delta \varphi) + i \sum_{n\in \mathbb{Z}} (-1)^{n}\text{sgn}(n) G^1(\Delta t-2n L_\lambda),\ee
where we define the $n=0$ term of the sum to be $G^1(\Delta t)$ --- although note that the Fourier transform of the $\tan$, as naively calculated above, does not contain this $n=0$ term. It is important to realize that (\ref{eq:naiveWHGR}) is merely a schematic expression: we did not treat the integration over $\omega$ carefully (by closing its contour and picking up appropriate residues of poles), and moreover the expression (\ref{eq:extBTZfinalprop}) we obtained using the hybrid WKB method is only approximatively valid.\footnote{In particular, it will have apparent poles in the complex frequency plane with non-zero imaginary part. However, 
the correct full propagator in frequency space will necessarily only have poles on the real axis, as the geometry only admits normal modes and no quasinormal modes.} In \cite{nextpaper}, we will use real time holography and carefully obtain a precise expression for an analogous correlator in the non-rotating BTZ background. Despite its shortcomings, (\ref{eq:naiveWHGR}) captures the most important behaviour: one can clearly see the appearance of ``echoes'' in the propagator: recurrences at periodic intervals $2L_\lambda$ of the correlator. Thus, at early times $0<\Delta t\ll 2 L_\lambda$, the propagator will mimick the BTZ propagator and decay exponentially; only at a time $\Delta t\sim 2 L_\lambda$ will the propagator start to differ qualitatively from the black hole one due to the appearance of the first echo.\footnote{For the extremal BTZ black hole, after the inital exponential falloff, the correlator transitions to a polynomial falloff due to the contribution of scattering events with period $2\pi$, originating from the compactness of the $\varphi$ coordinate, known as "images". See \cite{Bena:2019azk} for detailed discussion.} We show this behaviour in figure \ref{fig: one lambda}.

Even though the scalar two-point function in a single wormhole geometry has clear echoes, if we take such wormholes seriously as (toy models for) black hole microstates, a typical microstate will be a superposition of many such geometries. Then generically one expects the echo behaviour of such a typical state to be exponentially suppressed (in the black hole entropy). We discuss this in more detail in section \ref{sec:lambdaensemble}.

\section{Connecting BTZ to Flat Space}\label{sec:connectflat}
In the previous section, we calculated the two-point correlation function of a probe scalar field in the extremal BTZ black hole spacetime (\ref{eq:BTZmetric}), as well as its extremal Solodukhin wormhole counterpart (\ref{eq:WH3extmetric}). We saw how the wormhole gave rise to normal modes $\omega_n$ given by (\ref{eq:extBTZNMs}) in frequency space with spacing $\Delta\omega_n= \pi/L_\lambda$ (with the wormhole throat length $L_\lambda$ given by (\ref{eq:WH3extdefthroatlength}) for the extremal wormhole) so that the position space correlator correspondingly has echoes, as discussed in section \ref{sec:WHPSprop} .

In this section, we discuss how these asymptotic $\text{AdS}_3$ correlators can be related to (5D) flat space gravitational waves (as modeled by a scalar probe). This relation is given by the well-known decoupling limit of the five-dimensional asymptotically BMPV black hole \cite{Breckenridge:1996is}, or more precisely its three-charge generalization \cite{Breckenridge:1996sn}. Our focus will be on solving the minimally coupled probe scalar wave equation on this background (and its corresponding Solodukhin wormhole), relating the (quasi)normal modes of the asymptotically flat five-dimensional black hole (resp.\ wormhole) to the (quasi)normal modes of the extremal BTZ black hole (resp. extremal wormhole) through the decoupling limit. We will use methods similar to \cite{Chakrabarty:2019ujg,Chakrabarty:2015foa} (see also \cite{Taylor:1998tk,Dias:2007nj} for similar ideas and methods).

\begin{figure}[ht]
\centering
 \includegraphics{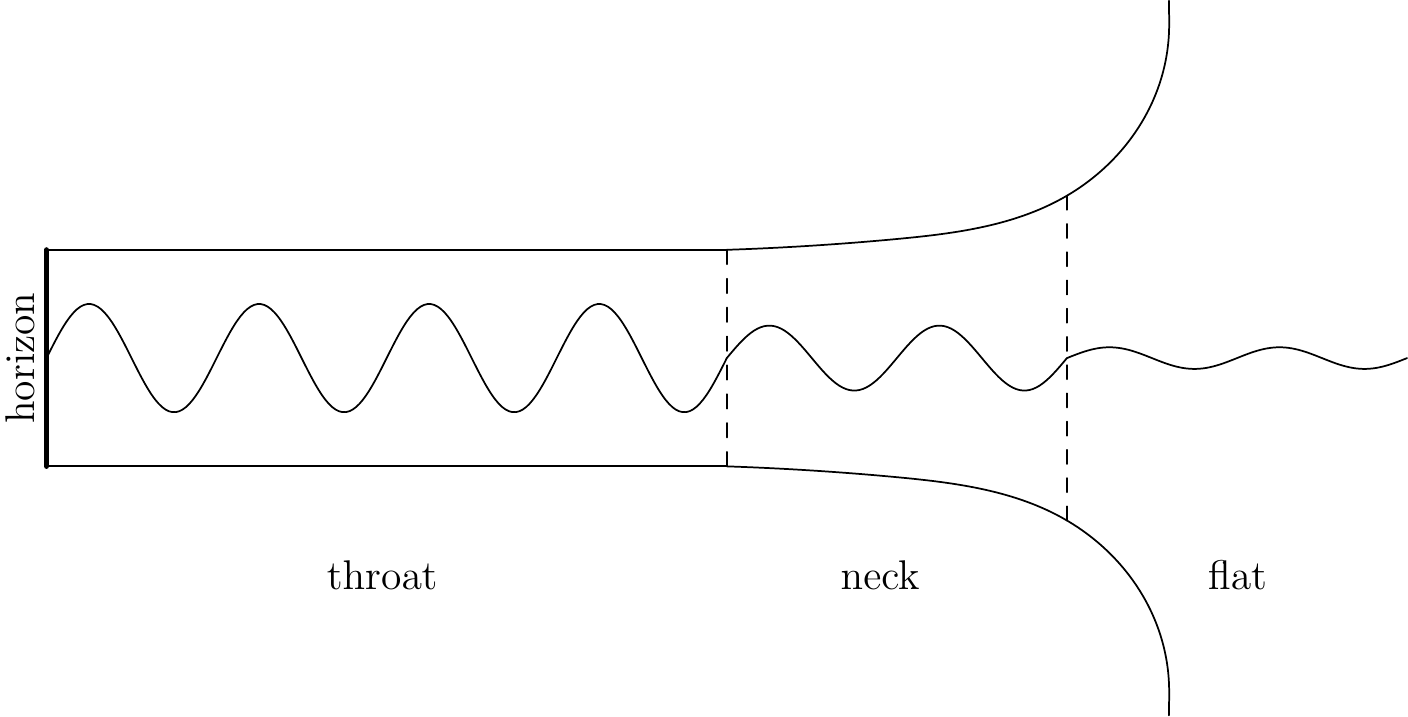}
 \caption{A schematic depiction of how the BTZ black hole is connected to flat space. In the decoupling limit $\epsilon\rightarrow 0$, the flat space region is decoupled from the BTZ region; when $1\gg\epsilon>0$, the flat space region is (weakly) coupled to the BTZ region through the intermediate neck region, which allows the scalar field to ``leak'' out of the BTZ region.}
 \label{fig:schemBTZflat}
\end{figure}

As we saw before (in section \ref{sec:AdS3WKB:ext}), the Solodukhin wormhole that corresponds to the extremal BTZ geometry has only normal modes:  the eigenfrequencies are real, with an approximate spacing between modes of $L_\lambda^{-1}$. We will see here (in section \ref{sec:flatQNMsWH}) that by coupling this wormhole to flat space, or said otherwise, by constructing the Solodukhin wormhole corresponding to the BMPV black hole, these normal modes receive a small imaginary part, making them quasinormal modes. This is related to the wave being able to ``leak out'' into flat infinity; see figure \ref{fig:schemBTZflat}. The real and imaginary parts of the resulting five-dimensional wormhole quasinormal modes follow the general scaling discussed in \cite{Cardoso:2019rvt} for (flat space) compact objects, as we discuss briefly in section \ref{sec:flatinterpretation}. Note that in principle the quasinormal modes are sufficient information to determine the entire analytic structure of the response function, so that they give a complete characterization of the system (see section \ref{sec:flatAdSdictionary}).

\subsection{Scalar Probe in BMPV Solution and the Decoupling Limit}
In this section, we will first describe the decoupling limit of BMPV, which gives the extremal BTZ black hole in $\text{AdS}_3$. Then, we describe the probe scalar wave equation in this background and solve it in three different regions where explicit analytic solutions are possible.

\subsubsection{BMPV and decoupling limit}
The three-charge asymptotically flat non-rotating BMPV black hole \cite{Breckenridge:1996sn} has as 6D metric:
\be \label{eq:6Dstartmetric} ds_{6D}^2=(H_1 H_5)^{-1/2} \left(-dt^2 + dy^2 + \frac{Q_P}{\rho^2}(dy-dt)^2\right) + (H_1 H_5)^{1/2} (d\rho^2 + \rho^2d\Omega_3^2),\ee
with:
\be H_1 = 1 + Q_1/\rho^2, \qquad H_5 = 1 + Q_5/\rho^2.\ee
The $y$-direction is compact with period $2\pi R_y$; $d\Omega_3^2$ is the volume form on the unit radius $S^3$ sphere. Thus, this solution can be seen as a six-dimensional black string, or after reduction along the compact $y$ direction, a 5D black hole. Note that the (finite size) horizon is at $\rho=0$.

This black hole can be uplifted to a full 10D type IIB supergravity solution \cite{Larsen:1999uk}, where then the parameters $Q_1,Q_5$ can be seen as the D1- and D5-brane charges of the solution, where both wrap the $y$ direction and the D5-branes also wrap the compact $M_4$; $Q_P$ is the momentum travelling along the $y$-direction. These charges are related to the quantized number of branes $N_1,N_5$ and quanta of momentum $N_P$ as: 
\be \label{eq:BMPVcharges} Q_1 = c_1 \frac{\alpha'^3 }{V_4} N_1, \qquad Q_5 = c_5 \alpha' N_5, \qquad Q_P = c_P \frac{\alpha'^4}{V_4R_y^2} N_P,\ee
where $\alpha'=l_s^2$ is the string length squared and $V_4=vol(M_4)$. The $c_I$ are factors which contain numerical factors as well as powers of the string coupling $g_s$, which will not enter in our discussion (see for example \cite{Maldacena:1996ky}).

In the (near-)decoupling limit of this metric, we realize that we can decouple excitations near the horizon and those very far away from the black hole. Concretely, we take $\alpha'=R_y^2 \eps^2$ for a dimensionless parameter $\eps$ which is sent to zero in the strict decoupling limit \cite{Aharony:1999ti,Larsen:1999uk}. (This limit is sometimes called the ``large $R_y$ limit'' \cite{Larsen:1999uk,Chakrabarty:2019ujg}, as we are taking $R_y$ to be much larger than the string scale.) We further take $\rho\sim \eps^2$ so that $\rho/\alpha'$ remains fixed. Note that we must have $V_4\sim\alpha'^2\sim \eps^4$. The charges scale as: $Q_1,Q_5\sim \eps^2$ and $Q_P\sim \eps^4$. The resulting metric in the $\epsilon\rightarrow 0$ limit is given by taking the initial asymptotically flat metric (\ref{eq:6Dstartmetric}) and ``dropping the $1$'s'' in the functions $H_{1,5}$, we get:
\be \label{eq:6Ddecoupled} ds_{BTZ\times S^3}^2 = \frac{\rho^2}{R^2} \left( -dt^2 + dy^2 + \frac{Q_P}{\rho^2} (dt-dy)^2\right) + \frac{R^2}{\rho^2}d\rho^2 + R^2d\Omega_3^2.\ee
The metric (\ref{eq:6Ddecoupled}) is the direct product of an $S^3$ with constant radius $R=(Q_1Q_5)^{1/4}$ and an $\text{AdS}_3$ factor with the same radius --- the extremal rotating BTZ black hole with mass, angular momentum and horizon radius:
\be
M=J=2Q_P/R^2,\qquad r_+ = \sqrt{Q_P}.
\ee
To compare with the BTZ metric in (\ref{eq:BTZmetric}), take $\rho^2 = r^2-r_+^2$ and $\varphi = y/R$.\footnote{Note that the periodicity of $y$ is different compared to what would be expected from this coordinate transformation.}

If we had considered the region $\rho\sim \eps^0$ instead, the leading order metric would have been pure 6D flat space (more precisely: $\mathbb{R}^{4,1}\times S^1(y)$):
\be \label{eq:6Dflat} ds_{\text{6D,flat}}^2 = -dt^2 + dy^2 + d\rho^2 + \rho^2 d\Omega_3^2.\ee
True to its name, the decoupling limit $\epsilon\rightarrow 0$ thus decouples excitations in the asymptotically flat region (\ref{eq:6Dflat}) from the near-horizon $\text{AdS}_3$ region (\ref{eq:6Ddecoupled}). In the near-decoupling limit,  when $\epsilon>0$ (but still $\epsilon\ll 1$), the asymptotic flat region (\ref{eq:6Dflat}) is weakly coupled to the AdS region (\ref{eq:6Ddecoupled}); this results in excitations in the AdS region being able to weakly ``leak out'' into the flat space region; see figure \ref{fig:schemBTZflat}. This is precisely the effect we will see when we calculate the quasinormal modes of the BMPV black hole (\ref{eq:6Dstartmetric}).

\paragraph{Three regions in the decoupling limit}
As we mentioned above, in the AdS region we have $\rho\sim \epsilon^2$ or $\rho/R\ll 1$, whereas in the asymptotic flat region $\rho\sim \epsilon^0$ or $\rho/R\gg 1$. However, (in analogy with \cite{Chakrabarty:2019ujg,Chakrabarty:2015foa}) we will instead introduce a second small, dimensionless parameter $\sigma\ll 1$ by hand, so that the AdS region of the decoupling limit is given by:
\be \label{eq:regionAdS} 0\leq \frac{\rho}{R} \leq \sigma,\ee
whereas the flat space region is given by:
\be \label{eq:regionflat} \frac{\rho}{R} \geq \frac{1}{\sigma}.\ee
These regions obviously \emph{do not} overlap, and so we cannot match a solution from AdS directly to flat space. To match solutions between the AdS and the flat region, we will need to pass through the so-called ``neck'' region, which is defined as:
\be \label{eq:regionneck} \sigma \leq \frac{\rho}{R} \leq \frac{1}{\sigma}.\ee

\subsubsection{Solving the scalar wave equation}
We will consider the minimally coupled probe scalar wave equation:
\be \label{eq:scalarwaveeq6D} \nabla^2_6 \Phi=0,\ee
on the full BMPV metric background (\ref{eq:6Dstartmetric}). We use the separation ansatz:\footnote{For definitiveness, we use the (flat) three-sphere metric $d\Omega_3^2 = d\theta^2 + \sin^2\theta d\phi_1^2 + \cos^2\theta d\phi_2^2$, where $\theta$ ranges between $0$ and $\pi/2$, and $\phi_{1,2}$ are periodic with period $2\pi$.}
\be \Phi(t,\rho,y,\Omega_3) = \exp\left(-i \omega t + i \tilde m y + i m_1 \phi_1 + i m_2 \phi_2\right)\Theta(\theta) g(\rho),\ee
Note that we can write $\tilde m = m/R_y$ with $m,m_1,m_2$ integers. We will be looking for $\omega \sim 1/R_y$, which corresponds to large wavelength, low energy modes in the decoupling limit. Note that if $\textrm{Im}(\omega)<0$, the mode is exponentially decaying. The angular function is a solution to:
\be \Theta'' +2\cot2\theta\, \Theta' - (m_1^2\csc^2\theta+m_2^2 \sec^2\theta)\, \Theta= -\Lambda\, \Theta,\ee
with $\Lambda = \ell(\ell+2)$ and $\ell$ an integer.

\paragraph{Radial equation}
It is convenient to use the dimensionless radial coordinate $\chi = R_y^2/\rho^2$ to get the radial scalar wave equation in the form:
\be \label{eq:fullradialscalarwave} \chi ^2 g''(\chi ) + V(\chi )g(\chi ) = 0,\ee
with potential:
\begin{align} \label{eq:fullpotentialvars}
 4 V(\chi ) &= \frac{R_y^2}{\chi } \left[\omega^2-\tilde m^2\right] + \left[ 1-\nu^2\right] +  \frac{\chi }{R_y^2}\left[\kappa(\omega^2-\tilde m^2)\right] + \frac{\chi ^2}{R_y^4} \left[\Xi^2 (\omega-\tilde m)^2\right],
\end{align}
where we have defined:
\begin{align}
 \nu^2 &= (\ell+1)^2 - (Q_1+Q_5)(\omega^2-\tilde m^2) - Q_P (\omega-\tilde m)^2 = (\ell+1)^2 + \mathcal{O}(\epsilon^2),\\
 \kappa &= R^4 + (\epsilon^6)(Q_1+Q_5)Q_P\frac{(\omega-\tilde m)^2}{\omega^2-\tilde m^2} = R^4(1 + \mathcal{O}(\epsilon^2)),\\
 \Xi &= R^2 Q_P^{1/2}.
\end{align}
Note that $\kappa,\Xi \sim \epsilon^4$.

\paragraph{Solutions in three regions}
Now, we solve the radial wave equation (\ref{eq:fullradialscalarwave}) in the three regions in the three regions (\ref{eq:regionAdS}), (\ref{eq:regionflat}), and (\ref{eq:regionneck}) separately (note that $\chi^2 \partial_\chi^2\sim (\epsilon\, \sigma)^0$):
\begin{itemize}
\item In the AdS region (\ref{eq:regionAdS}), for which we implement the strict decoupling limit $\rho\sim \epsilon\, \sigma$ so that $\chi\sim (\epsilon\, \sigma)^{-2}$; the potential (\ref{eq:fullpotentialvars}) reduces to:
\be \label{eq:potentialAdS} 4V(\chi) = \left( \left[ 1-\nu^2\right] + \frac{\chi}{R_y^2}\left[(\kappa(\omega^2-\tilde m^2)\right] + \frac{\chi^2}{R_y^4} \left[\Xi^2 (\omega-\tilde m)^2\right]\right) + \mathcal{O}(\epsilon^2\,\sigma^2) .\ee
This is precisely the radial potential of a scalar field in the 3D extremal BTZ metric (\ref{eq:BTZmetric}), where the scalar has the 3D mass $m^2 = \nu^2-1$ and  the effective extremal BTZ parameters $R, r_{+}$ are given by $R = \kappa^{1/4}, r_{+} = \Xi/\kappa^{1/2}$.
The solutions to the wave equation with the potential (\ref{eq:potentialAdS}) are given by Whittaker $W,M$ functions (which could also be expanded in terms of hypergeometric functions):
\begin{align}
 \nonumber g(\chi) = & c^A_1\, M\left(i\frac{\kappa}{4\Xi} (\omega+\tilde m), \frac{\nu}{2}, -i \chi \frac{\Xi}{R_y^2}(\omega-\tilde m)\right) \\
 \label{eq:solAdSregion} & + c^A_2\, W\left(i\frac{\kappa}{4\Xi} (\omega+\tilde m), \frac{\nu}{2}, -i \chi \frac{\Xi}{R_y^2}(\omega-\tilde m)\right),
 \end{align}
\item In the flat region, we have $\rho\sim \epsilon/\sigma$ so that $\chi\sim \sigma^2/\epsilon^2$, we find:
\be \label{eq:potentialflat} 4 V(\chi) = \left(\left[ 1-\nu^2\right]+(\frac{R_y^2}{\chi} \left[\omega^2-m^2\right]\right) + \mathcal{O}(\epsilon^2\sigma^2), \ee
The solutions to the wave equation with this potential are given by Bessel functions:
\be g(\rho)= \frac{1}{\rho}\left(c^F_1 I_{-\nu}(i \sqrt{\omega^2-m^2} \rho ) + c^F_2 I_{\nu}(i \sqrt{\omega^2-m^2} \rho )\right).\ee
The $I_{\nu}$ are the modified Bessel functions of the first kind.\footnote{We do not need to be careful with using Bessel functions of the first \emph{and} second kind since $\nu$ is not an integer at $\mathcal{O}(\epsilon^2)$. Thus, $I_\nu$ and $I_{-\nu}$ will not be linearly dependent.}
\item Finally, in the neck region, $\rho\sim \epsilon\, \sigma^0$ so that $\chi\sim \epsilon^{-2}\sigma^0$, we have:
\be \label{eq:potentialneck} 4V(\chi) = \left(\left[ 1-\nu^2\right] \right) + \mathcal{O}(\epsilon^2\sigma^0),\ee
The solution to the wave equation with this potential is given by a polynomial:
\be \label{eq:necksol} g(\rho) = c^N_1 \rho^{\nu-1} + c^N_2 \rho^{-\nu-1}.\ee
\end{itemize}

\paragraph{Matching solutions across the neck}
To obtain a full solution for the scalar wave, we must glue the different solutions together inside the neck region. 
We need to match the AdS region solutions $g(\rho)|_{\rm AdS}$ and the neck solution $g(\rho)|_{\rm neck}$ at a fiducial radius $\rho_A$ inside the neck,  $g(\rho_A)|_{\rm AdS} = g(\rho_A)|_{\rm neck}$ and $g'(\rho_A)|_{\rm AdS} = g'(\rho_A)|_{\rm neck}$. Similarly, we match the flat space solution to the neck solution at a different fiducial radius $\rho_F$.

The AdS solutions for large $\rho$ ( approaching the neck region), as well as the flat solutions for small $\rho$, both are of the form (\ref{eq:necksol}). We can define the following ratio of coefficients in the neck region:
\be \Upsilon_{\text{(region)}} \equiv\frac{c^{\text{(region)}}_1}{c^{\text{(region)}}_2},\ee
where (region) is either flat or AdS, and the subscript 1 (resp. 2) denotes the coefficient of the $\rho^{\nu-1}$ (resp. $\rho^{-\nu-1}$) piece (just as in (\ref{eq:necksol})).

In general, matching the solutions will be complicated and depend on the matching points $\rho_A,\rho_F$. However, it turns out that to $\mathcal{O}(\epsilon^2)$, the matching does not depend on $\rho_A,\rho_F$ and is precisely equivalent to simply setting the coefficients $c^{\text{(region)}}_i$ equal, which (apart from an overall normalization factor) amounts to the physical condition that the ratios defined above are equal:
\be \label{eq:ratiocondition} \Upsilon_{\rm AdS} = \Upsilon_\text{flat}\left(1+\mathcal{O}(\epsilon^2)\right).\ee

\paragraph{Quasinormal mode conditions}
To find a solution to the scalar wave equation everywhere in the BMPV spacetime, we need to solve the wave equation separately in the flat space and AdS regions and match these solutions across the neck region using (\ref{eq:ratiocondition}). We will focus on finding the quasinormal modes, which are special solutions to the wave equation where the following two conditions are imposed:
\begin{itemize}
 \item In the asymptotic flat region, we must demand that there are \emph{no incoming waves from spatial (flat) infinity}: as $\rho\rightarrow \infty$, the coeficient of $\exp\left(i \rho \sqrt{\omega^2-\tilde m^2}\right)$ must vanish in the flat space solution.
 \item In the AdS region, we must demand that there are \emph{no outgoing waves at the horizon}: as $\rho\rightarrow 0$, the coefficient of $\exp\left(-i\omega/\rho^2\right)$ must vanish in the AdS solution. 
\end{itemize}
The flat and AdS solutions both have two integration constants each, of which one is fixed by an overall normalization of the wave. The remaining two constants are fixed by the two conditions above. Then, the ratios $\Upsilon$ are given by:
\begin{align}
  \label{eq:ratioAdS} \Upsilon_{\rm AdS}& = \frac{(-i \Xi (\omega-\tilde m))^{-\nu} \Gamma(\nu) \Gamma\left(\left(\frac14(2-2\nu-i\frac{\kappa}{\Xi}(\omega+\tilde m)\right)\right)}{ \Gamma(-\nu)  \Gamma\left(\left(\frac14(2+2\nu-i\frac{\kappa}{\Xi}(\omega+\tilde m)\right)\right)},\\
  \label{eq:ratioflat} \Upsilon_\text{flat} &= \frac{4^{-\nu} e^{-2i\pi\nu}\nu (\tilde m^2-\omega^2)^{\nu}\Gamma(-\nu)}{\Gamma(1+\nu)}.
\end{align}
Thus, the matching condition (\ref{eq:ratiocondition}) imposes non-trivial constraints on the parameters of the solution, and in particular will only have solutions for particular, quantized value $\omega(\tilde m, \ell)$ --- the quasinormal modes of the BMPV black hole.

\subsection{Scalar Quasinormal Modes in BMPV}
As derived in the previous subsection, the condition to have a quasinormal mode scalar solution in the full BMPV black hole is (\ref{eq:ratiocondition}) such that $\Upsilon_{\rm AdS} = \Upsilon_\text{flat}$ with the ratios given by (\ref{eq:ratioAdS})-(\ref{eq:ratioflat}). We will now solve this equation to find the quasinormal modes $\omega(\tilde m, \ell)$ of the BMPV black hole.

\subsubsection{Leading order(s): BTZ quasinormal modes}
First, remembering that $\Xi\sim \epsilon^4$ and $\nu= (\ell+1)+\mathcal{O}(\epsilon^2)$, we can see that solutions to (\ref{eq:ratiocondition}) must be given by $\omega= \tilde\omega_n + \delta\tilde\omega_n$, where $\tilde\omega_n$ is a solution to $\Upsilon_{\rm AdS}(\tilde\omega_n)=0$, and $\delta\tilde\omega_n\sim \mathcal{O}(\epsilon^{4l})$.\footnote{One could be worried about $\ell=0$, as then naively $\tilde\omega_n$ and $\delta\tilde\omega_n$ would be of the same order. However, a simple calculation shows that the same reasoning holds for $\ell=0$, and in particular (\ref{eq:leadingorderomegan}) and (\ref{eq:deltaomegaBMPV}) are still valid for $\ell=0$. The reason that the same relations are still valid for $\ell=0$ is essentially that $\Upsilon_{\rm AdS}'(\tilde\omega_n)\sim \epsilon^{-4\ell-4}$ in (\ref{eq:corrgeneraldeltaomega}), which means $\delta\tilde\omega_n$ in (\ref{eq:deltaomegaBMPV}) is subleading to (\ref{eq:leadingorderomegan}), even for $\ell=0$.} 

Demanding $\Upsilon_{\rm AdS}(\tilde\omega_n)=0$ can only be achieved by frequencies $\tilde\omega_n$ that cause the $\Gamma$ function in the denominator to have a pole:
\be \label{eq:leadingorderomegan} \tilde\omega_n = -\tilde m - i 4\frac{\Xi}{\kappa}\qty(n + \frac{\nu+1}{2}) = -\tilde m -i 4 \frac{\sqrt{Q_P}}{R^2}\qty(n + 1 + \frac{\ell}{2}) + \mathcal{O}(\epsilon^2),\ee
for $n\geq 0$ integer. At $\mathcal{O}(\epsilon^0)$, these are precisely the quasinormal modes of extremal BTZ with $M=J=2Q_P/R^2$ and (3D) scalar mass $m_{\rm 3D}^2 =\ell(\ell+2)$ \cite{Chen:2010ik,Chen:2010sn} (as should be expected in the strict decoupling limit $\epsilon\rightarrow 0$). Using (\ref{eq:BMPVcharges}), we can see that these are indeed low-energy excitations with $\tilde\omega_n\sim R_y^{-1}$.

The expression (\ref{eq:leadingorderomegan}) are \emph{exact} zeros of $\Upsilon_{\rm AdS}$, which means the $\mathcal{O}(\epsilon^2)$ and higher order parts of this expression are the correct first corrections to the BTZ quasinormal modes due to the small but non-zero coupling $\epsilon$ to flat space; these remain the only $\epsilon$ corrections until the subleading corrections from $\Upsilon_\text{flat}$ contribute as well, which is when we must solve for $\delta\tilde\omega_n$.

\subsubsection{Subleading corrections from flat space}\label{sec:BMPVQNFsubleadingflat}
Since the quasinormal modes (\ref{eq:leadingorderomegan}) satisfy $\Upsilon_{\rm AdS}(\tilde\omega_n)=0$, the correction $\delta\tilde\omega_n$ that is necessary to solve (\ref{eq:ratiocondition}) must be:
\be \label{eq:ratioseps} \Upsilon_{\rm AdS}(\tilde\omega_n+\delta\tilde\omega_n) = \Upsilon_\text{flat}(\tilde\omega_n),\ee
or, solving for $\delta\tilde\omega_n$:
\be \label{eq:corrgeneraldeltaomega} \delta\tilde\omega_n = \frac{\Upsilon_\text{flat}(\tilde\omega_n)}{ \Upsilon_{\rm AdS}'(\tilde\omega_n)}.\ee

In $\Upsilon_{\rm AdS}'(\tilde\omega_n)$, the derivative must hit the $\Gamma$ function in the denominator of (\ref{eq:ratioAdS}) that is responsible for the vanishing of $\Upsilon_{\rm AdS}$. This gives:
\be \partial_{\omega}\left( \left[\Gamma\left(\left(\frac14(2+2\nu-i\frac{\kappa}{\Xi}(\omega+\tilde m)\right)\right)\right]^{-1}\right)_{\omega=\tilde\omega_n} = -i (-1)^{n} n! \frac{\kappa}{4\Xi}. \ee
The full expression for the correction (\ref{eq:corrgeneraldeltaomega}) is then given by:
\be \label{eq:deltaomegaBMPV} \delta\tilde\omega_n = (-1)^{\ell+1}\frac{ 2^{1-2 \ell} }{(\ell!)^3 }\binom{\ell+n+1}{\ell+1}\frac{Q_P^{1+\ell/2}R^{2l}}{ Q_1+Q_5} \tilde\omega_{n,i}^\ell(4\tilde m^2 + \tilde\omega_{n,i}^2)^{\ell+1/2}\left( \cos (2\ell+1)\alpha + i \sin(2\ell+1)\alpha\right) ,\ee
where we have written the original solution $\tilde\omega_n$ of (\ref{eq:leadingorderomegan}) as $\tilde\omega_n = -\tilde m - i \tilde\omega_{n,i}$, and we have defined:
\be \alpha = \tan^{-1}\left(\frac{\tilde\omega_{n,i}}{2\tilde m}\right).\ee
The full quasinormal modes of the BMPV black hole are then given by $\tilde\omega_n+\delta\tilde\omega_n$. The correction $\delta\tilde\omega_n$ in (\ref{eq:deltaomegaBMPV}) is clearly a $\mathcal{O}(\epsilon^{4\ell +2})$ correction to $\tilde\omega_n$ (analogous to the results in \cite{Chakrabarty:2015foa}). Both the real and imaginary part of the quasinormal frequencies get a (very) small correction due to the flat space coupling; when $\epsilon\ll1$, the full BMPV quasinormal modes are thus very well approximated by the BTZ quasinormal modes.

\subsection{Scalar (Quasi)normal Modes in BMPV Wormhole}
Now, we will turn the BMPV black hole into a Solodukhin wormhole by changing the extremal BTZ black hole of the near-horizon (decoupling) geometry into its corresponding Solodukhin wormhole as discussed in section \ref{sec:extDSWH}. Note that we will consider the wormhole parameter $\lambda$ to be much smaller than the decoupling limit parameter $\epsilon$:
\be \lambda \ll \epsilon,\ee
so that it makes sense to keep only leading pieces in $\lambda$ while going to higher orders in $\epsilon$, as we do when calculating the quasinormal modes in the flat space geometry. Physically, this also makes sense as one would expect the quantum correction at the horizon (governed by $\lambda$) to be much smaller than the non-extremality (controlled by $\epsilon$).

\subsubsection{Solutions in the BTZ Wormhole}
To find the quasinormal modes of the BMPV Solodukhin wormhole, we need to replace the effective decoupling limit extremal BTZ metric in the AdS region by the corresponding extremal wormhole, and solve the scalar wave equation on the wormhole background. The extremal BTZ metric we are considering is given by (\ref{eq:BTZmetric}) with effective BTZ parameters $R = \kappa^{1/4}$ and $r_{+}=r_{-} = \Xi/\kappa^{1/2}$. (Remember also that the ``Schwarzschild'' type radial coordinate $r$ used in (\ref{eq:BTZmetric}) is related to the radial coordinate $\rho$ as $\rho^2 = r^2 -r_{+}^2$.)

As discussed in section \ref{sec:scalarwave}, the general solution $g^{\text{WH}}(r)$ to the radial scalar wave equation for the wormhole geometry is given by the corresponding solution to the black hole wave equation, $g^{\text{WH}}(r)=g^{\text{BH}}(r)$, with the important condition that the wormhole radial solution and its first derivative is continuous at the wormhole throat radius $r_t$ given in (\ref{eq:rstarextWH}). Moreover, because the wormhole potential is even with respect to the wormhole throat position when written in the tortoise coordinate,  $V^{\text{WH}}(r_*)=V^{\text{WH}}(-r_*)$ (see (\ref{eq:fullWHpotential})), the normal mode solutions will be either even or odd themselves, which corresponds to the two possible conditions on the BTZ black hole solutions: 
\be \label{eq:WHbc} g_-^{BTZ}(r = r_t) =0, \qquad \partial_r g_+^{BTZ}(r = r_t) =0.\ee
 The conditions (\ref{eq:WHbc}) replace the BTZ black hole quasinormal condition that there are no ingoing waves at the horizon. The resulting scalar wave solution, expanded at large $\rho$ (or equivalently $r$), then gives rise to a new ratio $\Upsilon_\text{WH}$ (replacing (\ref{eq:ratioAdS})), given by:
\begin{align}\nonumber \Upsilon_\text{WH}(\omega) &= \Xi^{-\nu}\frac{(-i (\omega-\tilde m))^{-\nu} \Gamma(\nu)\Gamma(1+\nu) \Gamma\left(\frac14( 2-2\nu-i(M+W))\right)}{\Gamma\left(\frac14( 2+2\nu-i(M+W))\right)} \\
 \nonumber & \times X_\pm\Bigg[\Gamma(-\nu)\Gamma(1+\nu)X_\pm - (-i(\omega-\tilde m))^{i/2(W+M)}\Gamma\left(\frac14( 2-2\nu-i(M+W))\right)\\
  \label{eq:AdScoeffWH}& \qquad \quad \times\Gamma\left(\frac14( 2+2\nu-i(M+W))\right) \Bigg]^{-1},\\
 X_{\pm} &= \left(\frac{\Xi}{\kappa}\right)^{i/2(W+M)}\Bigg( \mp e^{\frac{-2i(W-M)}{\lambda^2}} \left(\frac{\lambda^2}{2}\right)^{i(W+M)/2}\nonumber\\
 & - i e^{-i/2\pi\nu}  (W-M)^{i/2(M+W)}\frac{\Gamma\left(\frac14( 2+2\nu-i(M+W))\right)}{\Gamma\left(\frac14( 2+2\nu+i(M+W))\right)}\Bigg),\\
 \label{eq:rescaledMW} M &= \frac{\kappa}{\Xi} \tilde m, \qquad W =\frac{\kappa}{\Xi} \omega,
\end{align}
where $X_{-}$ (resp. $X_+$) corresponds to the odd (resp. even) modes.

\subsubsection{Leading order(s): wormhole normal modes}
To find the quasinormal modes of the BMPV wormhole, we now have to solve the altered version of (\ref{eq:ratiocondition}):
\be \label{eq:ratioconditionWH} \Upsilon_\text{WH}(\omega) = \Upsilon_\text{flat}(\omega)\left(1+\mathcal{O}(\epsilon^2)\right),\ee
with $\Upsilon_\text{WH}(\omega)$ given by (\ref{eq:AdScoeffWH}) and $\Upsilon_\text{flat}(\omega)$ still given by (\ref{eq:ratioflat}). A similar scaling analysis as equation (\ref{eq:ratioseps}) for (\ref{eq:ratioconditionWH}) applies, and in particular the leading order solutions to (\ref{eq:ratioconditionWH}) are given by $\Upsilon_\text{WH}(\omega_n)=0$. This is equivalent to demanding $X_\pm=0$.\footnote{Although the quasinormal modes $\tilde \omega_n$ in (\ref{eq:leadingorderomegan}) of the BTZ black hole also satisfy $\Upsilon_\text{WH} = 0$, these will not correspond to actual (quasi)normal modes of the wormhole, as a complete analysis of the allowed solutions in the wormhole would reveal.}

This transcendental equation cannot be analytically solved in general. In the large $\omega$ limit, we find the approximate solutions:
\be \omega_n = \tilde m  + \left(n+\ell+\frac12\right)\frac{\pi}{L_\lambda} + \mathcal{O}(L_\lambda^{-2}, \tilde m L_\lambda^{-1} \log L_\lambda),\ee
where we used (\ref{eq:WH3extdefthroatlength}),
where $n$ is an integer (to $\mathcal{O}(\epsilon^2)$); when $n$ is even (resp. odd), the normal mode corresponds to an even (resp. odd) mode obtained by $X_+=0$ (resp. $X_-=0$). These high frequency normal modes (up to the $\mathcal{O}(\tilde m L_\lambda^{-1}\log L_\lambda)$ term) agree with  what we found before in (\ref{eq:extBTZNMs}) with the substitution\footnote{This substitution is necessary since $y$ in (\ref{eq:6Ddecoupled}) is periodic with $2\pi R_y$ and $\varphi$ is periodic with $2\pi$.} $k\rightarrow m R/R_y = \tilde m R$.

We can also solve (\ref{eq:ratioconditionWH}) in a low-frequency limit $\omega-\tilde m \sim L_\lambda^{-1}$:
\be \omega_n = \tilde m + \left(n+\frac{\ell}{2}\right)\frac{\pi}{L_\lambda}  +  \mathcal{O}(\tilde m L_\lambda^{-1}),\ee
where $n$ even (resp. odd) corresponds to the odd (resp. even) mode obtained by $X_-=0$ (resp. $X_+=0$).

\subsubsection{Subleading corrections from flat space}\label{sec:flatQNMsWH}
Now, we can calculate the corrections to the modes due to the coupling to flat space. We need to calculate $\Upsilon_{\rm AdS}'(\omega_n)$; this derivative must hit the $X_\pm$ in the numerator. Even though we do not have an analytic expression for the modes $\omega_{n}$, we can still simplify the resulting derivative and write the expression for the correction $\delta\omega_n$ explicitly in terms of $\omega_n$. The leading real and imaginary parts of $\delta\omega_n$ are then:
\begin{align}
   \delta\omega_n &= \frac{\Upsilon_\text{flat}(\omega_n)}{ \Upsilon_\text{WH}'(\omega_n)}\\
\nonumber  &= -\frac{1}{L_\lambda}\left(R^{2\ell+2}Q_P^{(1+\ell)/2}\right) \frac{(\omega_n^2-\tilde m^2)^{\ell} (\omega_n-\tilde m)^{\ell+1}}{2^{2\ell+1}\ell!^3(\ell+1)!}
\left|\Gamma\left(1+\frac{\ell}{2}+\frac{i}{4}(M+W_n)\right)\right|^2\\
\label{eq:deltaomegaBMPVWH}& \times  e^{\pi/4(M+W_n)}\left( \frac{1}{Q_1+Q_5} + i \frac{\pi}{2} \frac{\omega_n^2-\tilde m^2}{\ell+1}   \right).  \end{align}
The real part is a correction to the quasinormal frequency at order $\mathcal{O}(\epsilon^{4\ell+2})$ (as it was for the BMPV black hole frequencies in section \ref{sec:BMPVQNFsubleadingflat}) whereas the imaginary part enters at $\mathcal{O}(\epsilon^{4\ell+4})$ (similar to \cite{Chakrabarty:2015foa}). Of course, the correction to the real part is small compared to the original (real) normal mode $\omega_n$. Note that\footnote{Although we do not show it explicitly, we have $\omega\geq \tilde m$ for all normal modes of the wormhole.} $\textrm{Im}(\delta\omega_n)<0$, as appropriate for a decaying mode.

\subsection{Interpretation}\label{sec:flatinterpretation}
In this section, we calculated the quasinormal modes for the BMPV black hole, as well as for the Solodukhin wormhole modification of BMPV. For both examples, we solved the wave equation in the near-decoupling limit (with small decoupling parameter $\eps$), and found that near-horizon, asymptotic AdS (decoupling) region gives the most important contribution to the quasinormal modes. As we will discuss in the next section \ref{sec:flatAdSdictionary}, the quasinormal modes of a system determine the analytic structure of the relevant response function; thus, they capture the physics of the reflection and transmission coefficients as well as the entire correlator.

For the BMPV black hole, the (small) coupling to flat space provides a small correction to both the real and imaginary parts which is subleading in $\eps$. For the BMPV wormhole, the key difference is that {the} wormhole itself has only non-decaying normal modes, and the imaginary part of the full quasinormal mode is thus entirely due to the (small) coupling to flat space. Physically, this coupling to flat space allows the scalar wave to ``leak out'' at (flat) infinity instead of being trapped in the asymptotically AdS wormhole ``box'' forever (see figure \ref{fig:schemBTZflat}). Thus, the ``echoes'' that a scalar field in the asymptotically flat BMPV black hole will have are determined by the real part of the quasinormal modes and hence  by the behaviour of the scalar field in the AdS spacetime. The small coupling to flat space and the resulting small imaginary part of the quasinormal mode simply gives a small attenuation of the echo amplitude over time; this attenuation is directly determined by the travel time down the BTZ throat $2L_\lambda$. So, we can conclude that the relation between the behaviour of the scalar field signal (in position space) in the BMPV black hole and BMPV wormhole is captured by the relation between the BTZ black hole and BTZ wormhole behaviours of the scalar field, and thus by the holographic correlators for those solutions that we studied in section \ref{sec:AdS3WKB}.

From the CFT perspective, the normal modes of the wormhole correspond to the poles of the scalar two-point correlator. The ``leaking'' into flat space can be thought of as the result of perturbing the CFT with an irrelevant deformation \cite{Avery:2009tu} which changes the UV and thus also perturbes the scalar correlator. As long as the irrelevant deformation is small (as it is here, controlled by $\epsilon\ll 1$), the deformation can be treated perturbatively in the CFT \cite{Baggio:2012db}.

Note that the imaginary part of the quasinormal frequencies is parametrically small compared to the real part. As we saw above in (\ref{eq:deltaomegaBMPVWH}), the imaginary part of the quasinormal modes is $\mathcal{O}(\epsilon^{4\ell+4}\lambda^2)$: it is both small due to the small part of the scalar wave escaping to infinity (the $\epsilon$ dependence), as well as parametrically small in the wormhole parameter $\lambda$.

The resulting system of quasinormal frequencies for the BMPV wormhole is reminiscent of other flat space Solodukhin type wormholes \cite{Bueno:2017hyj,Barack:2018yly}. In particular, we can compare the scaling of real and imaginary parts of the quasinormal modes with the analysis of sec 4.1 in \cite{Cardoso:2019rvt} for the quasinormal modes of general wormholes (or other ``cavity'' potentials) in (4D) flat space. Defining $\hat{\omega}_n = \omega_n - \tilde m$, we see that:
\be \hat{\omega}_n \sim \frac{1}{L_\lambda}.\ee
The real part of the quasinormal mode scales as the inverse ``cavity size'' (in tortoise coordinates). Moreover, we can also see from (\ref{eq:deltaomegaBMPVWH}) that the imaginary part scales as:
\be \label{eq:scalingImomega} \textrm{Im}(\delta\omega_n) \sim \frac{1}{L_\lambda}\hat{\omega}^{2\ell+2} \sim \hat{\omega}^{2\ell+3},\ee
in precise agreement with the scaling found in eq. (34) of \cite{Cardoso:2019rvt}. This scaling of the imaginary part of the mode can be understood as the product of the tunneling amplitude $|\mathcal{A}|^2\sim \hat{\omega}_n^{2\ell+2}$ of the wave \cite{Starobinskil:1974nkd} to reach outside the cavity, with the timescale $\Delta t\sim L_\lambda^{-1}$ needed for the wave to travel once through the cavity of length $L_\lambda$ \cite{Cardoso:2019rvt}. Note that the general scaling (\ref{eq:scalingImomega}) also holds in the asymptotically flat microstate geometries considered in \cite{Bena:2020yii}; there, the ``cavity size'' is determined by the (large) redshift down the microstate throat, $n_1n_5/j_L$.

 \section{Relation to Echo Literature in Flat Space}\label{sec:flatAdSdictionary}

In this section, we make the connection to wave scattering in flat space. To keep the discussion general, we now consider the Schr\"odinger problem with source $S(x,\omega)$:
\begin{equation}
\frac{d^2 \psi(x)}{d x^2} - V(x) \psi (x) = S(x,\omega)\,,\label{eq:Schrodingerflat}
\end{equation}
where $V$ is a potential and  $S$ is a source. This equation can describe, for example, the scattering problem of a gravitational wave in Fourier space (where $x$ is the tortoise coordinate); or it can be used to find quasinormal modes when the source vanishes.

We will recall first how to obtain the solution with a source $S(x)$ in flat space and what the precise relation is to the quasinormal modes and echoes, and then proceed to generalize the discussion to include AdS correlators and quasinormal modes discussed above.

Although we focus on four-dimensional black holes in section \ref{sec:wavesflat}, the results apply to general spacetime dimensions $d$, unless specified otherwise. In different dimensions, the form of the potential may change, but its asymptotic behaviour remains similar.

\subsection{Waves in Flat Space}\label{sec:wavesflat}

\begin{figure}[!ht]
	\begin{subfigure}{.5\textwidth}
		\centering
		\includegraphics[width=0.9\linewidth]{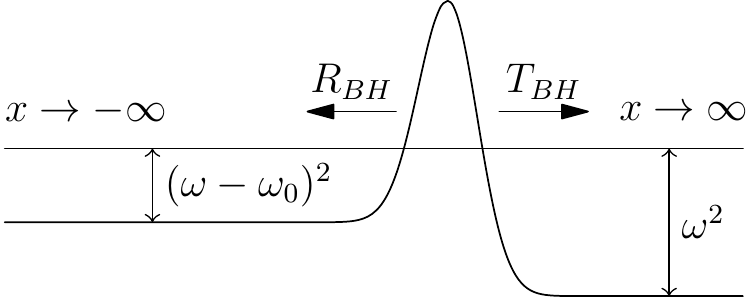}  
		\caption{}
		\label{fig: flat BH}
	\end{subfigure}
	\begin{subfigure}{.5\textwidth}
		\centering
		\includegraphics[width=0.9\linewidth]{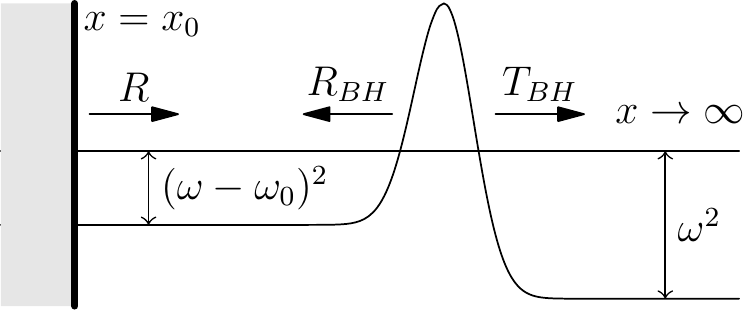}  
		\caption{}
		\label{fig: flat ECO}
	\end{subfigure}
	\caption{(a) A schematic plot of the radial potential of a rotating black hole with flat asymptotics. Asymptotic infinity is at $x\rightarrow \infty$, the horizon at $x \rightarrow -\infty$, and the bump is associated with the light-ring; (b) A schematic plot of the radial potential of a rotating ECO with flat asymptotics. Asymptotic infinity is at $x \rightarrow \infty$, and at large negative $x_0$ we put a wall which has arbitrarily complicated behaviour captured by the function $R(\omega,x_0)$.}
	\label{fig: flat potentials}
\end{figure}

This section is based on  the discussion of \cite{Mark:2017dnq}, and follows conventions of \cite{Cardoso:2019rvt}. For more details, we refer to those references.
For the scattering problem with a rotating black hole in four dimensions, the potential asymptotically approaches a negative constant, and all derivatives are zero at the asymptotic boundaries:
\begin{equation}
\lim_{x\to \infty} V(x) = -\omega^2\,,\qquad \lim_{x\to -\infty} V(x) = -(\omega-\omega_0)^2\,.
\end{equation}
In the above equation, $\omega$ is the frequency of the particular mode and $\omega_0$ a constant depending on the spin of the black hole. The potential has a maximum at the position of the light-ring, the location of the unstable photon orbit (for Schwarzschild in usual radial coordinates, this is at $r = 3M$). We give a schematic representation in figure \ref{fig: flat BH}; for the detailed form of potentials depending on the spin of the perturbation, see \cite{Zerilli:1971wd,Berti:2009kk}.

The wave equation has two homogeneous independent solutions, which we can choose to have the following asymptotic behaviour:
\begin{align}
\begin{aligned}
\text{ingoing at horizon:}& & \psi_- &\sim 
\begin{cases}
e^{-i (\omega-\omega_0) x}, &x\to -\infty\\
A_{\rm in}(\omega) e^{-i \omega x} + A_{\rm out}(\omega) e^{i \omega x}, &x\to +\infty\\
\end{cases} \\
\text{outgoing at infinity:}& & \psi_+ &\sim 
\begin{cases}
B_{\rm in}(\omega) e^{-i (\omega-\omega_0) x} + B_{\rm out}(\omega) e^{i (\omega-\omega_0) x}, &x\to -\infty\\
e^{i \omega x}, &x\to +\infty.
\end{cases}
\end{aligned}\label{eq:psipmflat}
\end{align}
One can find the transmission and reflection coefficients for waves coming in from infinity and scattering off the potential. It is most convenient for the current discussion  to define those coefficients for waves coming from the left, see figure \ref{fig: flat BH},
\begin{equation}
R_{\rm BH}(\omega) = \frac{B_{\rm in}(\omega)}{B_{\rm out}(\omega)}\,,\qquad T_{\rm BH}(\omega) = \frac 1 {B_{\rm out}(\omega)}\,.\label{eq:transmissionreflection+flat}
\end{equation}

The physical solution to the wave equation is composed only of ingoing waves at the horizon and outgoing waves at infinity
\begin{align}
\psi_{\text{BH}} \sim \begin{cases}
e^{-i (\omega-\omega_0) x}, &x\to -\infty\\
e^{i \omega x}, &x\to +\infty.
\end{cases}
\label{eq: bh bdy conds}
\end{align}
The black hole Green's function solves the wave equation with source $S = \delta (x-x')$ and obeys the boundary conditions (\ref{eq: bh bdy conds}):
\begin{align}
G_{\text{BH}}(x,x') = { \psi_+(\text{max}(x,x'))\psi_-(\text{min}(x,x'))\over W(\omega)},
\end{align}
where $W(\omega) = \psi_- \psi_+' -\psi_+ \psi_-'$ is the Wronskian. 
For this simple second order differential equation, the Wronskian is a constant and can be evaluated at infinity: $W(\omega)= 2 i \omega A_{\rm in}(\omega)$. 
The solution to the full wave equation is then
\begin{equation}
\psi_{\text{BH}} = \psi_+ \int^x_{-\infty}{S \psi_- \over W} + \psi_- \int^\infty_{x}{S \psi_+ \over W}\,.
\end{equation}
The frequencies of the QNMs, the solutions for $S=0$, are determined by the poles of the Green's function, or alternatively: the zeroes of the Wronskian $W(\omega_n)=0$. Note that these are also precisely the frequencies for which $\psi_+$ and $\psi_-$ are linearly dependent.

Now imagine that we have an exotic compact object (ECO), whose potential agrees very well with the black hole potential up until some very large negative Schr\"odinger coordinate $x_0\ll 0$, which is assumed to be located somewhere in the flat region of the potential, away from any possible light-ring structure. For $x<x_0$ the potential starts to significantly deviate from the black hole potential. One can model this behaviour by putting some generic $\omega$-dependent effective reflective boundary condition at $x_0$, see figure \ref{fig: flat ECO}. This changes the boundary condition for a purely outgoing mode at infinity to
\begin{align}
\psi_{\text{ECO}} \sim \begin{cases}
e^{-i (\omega-\omega_0) x}+ R(\omega, x_0)e^{i (\omega-\omega_0) x}, &x\to x_0\\
e^{i \omega x}, &x\to \infty,
\end{cases}
\label{eq: eco bdy conds bh}
\end{align}
where the reflection coefficient  $R(\omega,x_0)$ is determined by the details of the ECO model. Note that we follow \cite{Cardoso:2019nis} and have absorbed a factor of $e^{2 i \omega x_0}$ into $R$ with respect to \cite{Mark:2017dnq}.

We can write
\begin{equation}
\psi_{\rm ECO} = \psi_{\rm BH} + \psi_{\rm H}\,, 
\end{equation}
with $\psi_{\rm H}$ a homogenous solution such that $\psi_{\rm WH}$ obeys the boundary conditions (\ref{eq: eco bdy conds bh}). This means that $\psi_{\rm H} = A(\omega, x_0) \psi_+$ for some function $A$. Following \cite{Mark:2017dnq}, we take a judicious choice of multiplication constant for later comparison with the black hole solution as $A(\omega, x_0) = \mathcal{K}(\omega, x_0) \int_{-\infty}^\infty \frac{S \psi_+}W dx$:
\begin{align}
\psi_{\text{H}} &= \mathcal{K} \psi_+ \int^\infty_{-\infty}{S \psi_+ \over W}, 
\end{align}
 where $\mathcal{K}(\omega, x_0)$ is a constant called the `transfer function'
As $x \to x_0$, we have $\psi_{\rm BH} = \psi_+\int^\infty_{x_0}{S \psi_+ \over W}$ and $\psi_{\rm H} =  \mathcal{K}\psi_+ \int^\infty_{x_0}{S \psi_+ \over W}$, so that at large negative $x$ we find
\begin{align}
\psi_{\rm ECO} &= \left(e^{-i(\omega-\omega_0) x} + {\cal K}(\omega, x_0) (B_{\rm in}(\omega) e^{-i (\omega-\omega_0) x} + B_{\rm out}(\omega) e^{i(\omega-\omega_0) x}) \right) \int_{x_0}^\infty \frac{S \psi_+}W dx , 
\end{align}
Comparing with (\ref{eq: eco bdy conds bh}) gives:
\begin{equation}
{\cal K}(\omega) = \frac{{T}_{\rm BH}(\omega) {R}(\omega, x_0)}{1- {R}_{\rm BH}(\omega) {R}(\omega,x_0)}\,.\label{eq:definitionK}
\end{equation}
At large distances, relevant to a detector, the wave function has the form
\begin{align}
\psi_{\rm ECO}(x\to \infty) 
&=\, \psi_{\rm BH}(x\to \infty)
+ {\cal K}(\omega, x_0)e^{i(\omega + (\omega-\omega_0))x} \psi_{\rm BH}(x\to -\infty)\,,
\end{align}
hence the response of the ECO corresponds to the initial BH response, followed by an additional term that is determined by the reflectivity of the compact object. It is this second term that can be written as a train of echoes following the main signal, from the geometric series expansion:
\begin{equation}
{\cal K}(\omega, x_0)= {T}_{\rm BH}(\omega){R}(\omega, x_0)
\sum_{n=1}^{\infty} ({R}_{\rm BH}(\omega) {R}(\omega, x_0))^{n-1}.
\end{equation}

The ECO QNM frequencies $\omega_n$ are the poles of $\cal K$, or equivalently:
\begin{equation}
1- {R}_{\rm BH}(\omega_n) {R}(\omega_n,x_0) = 0,
\end{equation}
If we consider the simple example of a (possibly partially reflective) $\omega$-independent ``brick wall" located at $x_0$ and assume a slowly varying $R_{\rm BH}(\omega)$, we have
\begin{align}
R_{\rm BH}(\omega) \approx R_{BH}, \qquad R(\omega,x_0) = Re^{-2i(\omega-\omega_0) x_0}.
\end{align}
In this case, the asymptotic detector receives the initial signal from the ECO which decays very similarly to the black hole; then, after a time $\Delta t = 2 x_0$, a series of echoes follows. The QNM frequencies are approximately
\begin{equation}
\omega_ n = \omega_0 + \frac{\pi}{x_0} \left(n + {i \over 2 \pi} \ln( {R}_{\rm BH}R)\right) + \mathcal{O}\qty({\pi n/x_0 \over R_{\rm BH}(\omega_n)/ R_{\rm BH}'(\omega_n)}), \qquad n=0,1,2,\dots \ .
\end{equation}
The relation to Green's function, and in particular how a very similar (early-time) behaviour in the ECO and black hole backgrounds nevertheless is associated with vastly different QNM spectrums, was worked out in \cite{Hui:2019aox}.

\subsection{Waves in AdS}\label{sec:wavesAdS}

\begin{figure}[!ht]
	\begin{subfigure}{.5\textwidth}
		\centering
		\includegraphics[width=.9\linewidth]{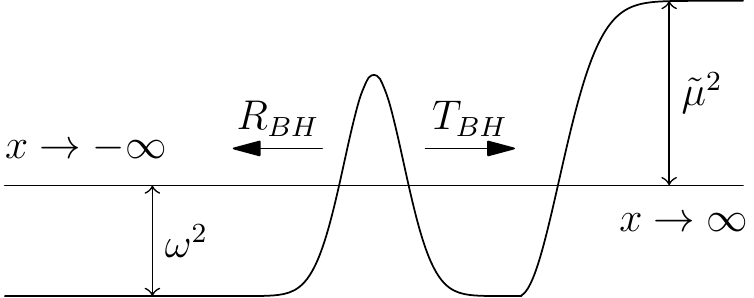}  
		\caption{}
		\label{fig: AdS BH}
	\end{subfigure}
	\begin{subfigure}{.5\textwidth}
		\centering
		\includegraphics[width=.9\linewidth]{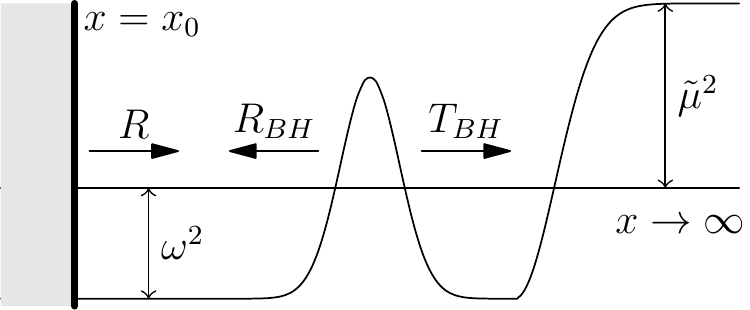}  
		\caption{}
		\label{fig: AdS ECO}
	\end{subfigure}
	\caption{(a) A schematic plot of the radial potential for a non-rotating black hole with AdS asymptotics. The boundary is at $x \rightarrow \infty$, the horizon at $x \rightarrow -\infty$. The bump is, similarily to flat space, associated with the light-ring; (b) A schematic plot of the radial potential of a non-rotating ECO with AdS asymptotics. The boundary is at $x \rightarrow \infty$, and at large negative $x_0$ we put a wall which has arbitrarily complicated behaviour captured by the function $R(\omega,x_0)$.}
	\label{fig: AdS potentials}
\end{figure}

Changing the asymptotics from flat space to AdS does not change the Schr\"odinger form of the wave equation \eqref{eq:Schrodingerflat}. We choose $x\to \infty$ to correspond to the asymptotic AdS boundary and $x \to -\infty$ to the horizon. For scattering problems in  black hole backgrounds in AdS, the potential asymptotically approaches a positive constant at the AdS boundary and a negative constant near the horizon:\footnote{For example, we can use a (rescaled) version of the WKB coordinates defined in section \ref{sec:AdS3WKB} for the BTZ black hole. Then we have $\tilde \mu^2 = (r_+^2/R^4) \mu^2$.}
\begin{equation}
\lim_{x\to \infty} V(x) = \tilde \mu^2\,, \qquad \lim_{x\to - \infty} V(x) \to -\omega^2\,,\qquad
\end{equation}
and all derivatives are zero at the asymptotic boundary. For simplicity, here we are taking non-rotating black holes (or ECOs) in AdS. Then, the potential asymptotes to the frequency of the mode $\omega$ at $x\rightarrow -\infty$.
The potential for a non-rotating AdS black hole is schematically depicted in figure \ref{fig: AdS BH}.

For easy comparison to \cite{Bena:2019azk} we take the two independent solutions to be:
\begin{align}
\begin{aligned}
\psi^{\rm grow}&\sim
\begin{cases}
B_{\rm in}^{\rm grow}(\omega) e^{-i\omega x}+B_{\rm out}^{\rm grow}(\omega) e^{i\omega x}, \qquad& x\to - \infty\cr
e^{\tilde \mu x}, \qquad &x\to \infty,
\end{cases}\\
\psi^{\rm dec} & \sim
\begin{cases}
B_{\rm in}^{\rm dec}(\omega) e^{-i\omega x}+B_{\rm out}^{\rm dec}(\omega) e^{i\omega x}, \qquad &x\to - \infty\cr
e^{-\tilde \mu x},\qquad &x\to \infty.
\end{cases}
\end{aligned} \label{eq:AdSgrowdec}
\end{align}
What is important to realize is that, although the boundary conditions (\ref{eq:AdSgrowdec}) are analogous to the flat space case (\ref{eq:psipmflat}), what is different in AdS is that the $x\rightarrow\infty$ boundary conditions imposed in (\ref{eq:AdSgrowdec}) immediately imply that $\psi^{\rm dec},\psi^{\rm grow}$ are \emph{real}\footnote{This follows because the real and imaginary parts of the solution must each satisfy the same Schr\"odinger equation as long as the potential $V(x)$ is real.} for all $x$. In particular, we have:
\be \label{eq:AdScc-cond} B_{\rm in}^{\rm grow} = \left(B_{\rm out}^{\rm grow}\right)^*, \qquad B_{\rm in}^{\rm dec} = \left(B_{\rm out}^{\rm dec}\right)^*.\ee

For AdS spacetimes, we will not introduce a bulk source $S(x,\omega)$; however, as we will see below, the presence of the growing mode is interpreted as a holographic source on the boundary using the AdS/CFT dictionary. 
We first discuss the QNMs. The appropriate boundary conditions for black hole QNMs are
\begin{equation}
\psi^{\text{non-sourced}}_{\rm BH}  \sim
\begin{cases}
e^{-i \omega x}, &x\to -\infty\\
e^{-\tilde \mu x}, &x\to +\infty.
\end{cases}
\end{equation}
From the decaying mode, we can again define reflection and transmission coefficients, for waves coming from the left, see figure \ref{fig: AdS BH},
\begin{equation}
{R } _{\rm BH}(\omega) = \frac{B_{\rm in}^{\rm dec}(\omega)}{B_{\rm out}^{\rm dec}(\omega)}, \qquad {T}_{\rm BH}(\omega) = {1 \over B_{\rm out}^{\rm dec}(\omega)}.
\end{equation}
The frequencies of the black hole QNMs are then precisely given by the poles of ${R}_{\rm BH}(\omega)$. Note that from (\ref{eq:AdScc-cond}), it follows that:
\be \label{eq:AdSRTrel} |{R }_{\rm BH}| = 1  ,\qquad {T}_{\rm BH} = \abs{T_{\rm BH}} ({R }_{\rm BH})^{1/2},\ee

A solution at a generic frequency will include the growing mode (see \eqref{eq:alphabeta_first}):
\begin{equation}
\psi^{\rm sourced}_{\rm BH}  = \beta_{\rm BH}(\omega) \psi^{\rm grow} + \alpha_{\rm BH}(\omega) \psi^{\rm dec}, \qquad \beta_{\rm BH} B_{\rm out}^{\rm grow} + \alpha_{\rm BH} B_{\rm out}^{\rm dec} =0,
\end{equation}
where the condition on the coefficients is there to ensure only ingoing modes at the horizon. The coefficient $\beta_{\rm BH}$ is proportional to the holographic source of the CFT boundary operator, and $\alpha_{\rm BH}$ is proportional to the vacuum expectation value of the boundary operator. The holographic CFT (retarded) two-point function for this boundary scalar operator is then given by:
\begin{equation}\label{eq:AdScorrT}
\mathcal{R}_{\rm BH}(\omega) = \frac{\alpha_{\rm BH}(\omega)}{\beta_{\rm BH}(\omega)}  = - {T}_{\rm BH}(\omega) B_{\rm out}^{\rm grow}(\omega)\,.
\end{equation}

If we now take an ECO, generically the solution will instead be (valid for all $x$): 
\begin{align}
\psi^{\rm sourced}_{\rm ECO} = \beta_{\rm ECO}(\omega) \psi^{\rm grow} + \alpha_{\rm ECO}(\omega) \psi^{\rm dec}.
\label{eq: eco in terms of grow and dec}
\end{align}
Just like in flat space, we can assume that the potential of the ECO is described by the black hole one up until a point $x_0\ll 0$, deep in the IR, where it starts to deviate significantly. We can again model this as an $\omega$-dependent reflection at large negative $x_0$, see figure \ref{fig: AdS ECO}. This changes the IR boundary conditions as
\begin{align}
\psi^{\rm sourced}_{\text{ECO}} \sim \begin{cases}
e^{-i \omega x}+ R(\omega, x_0)e^{i \omega x}, &x\to x_0\\
\beta_{\rm ECO}(\omega) \psi^{\rm grow} + \alpha_{\rm ECO}(\omega) \psi^{\rm dec}, &x\to \infty.
\end{cases}
\label{eq: eco bdy conds}
\end{align}
Since (\ref{eq: eco in terms of grow and dec}) is valid everywhere, the boundary conditions (\ref{eq: eco bdy conds}) give the relation:
\begin{equation}
\frac{\beta_{\rm ECO}(\omega) B_{\rm out}^{\rm grow}(\omega) + \alpha_{\rm ECO}(\omega) B_{\rm out}^{\rm dec}(\omega)}{\beta_{\rm ECO}(\omega) B_{\rm in}^{\rm grow}(\omega) + \alpha_{\rm ECO}(\omega) B_{\rm in}^{\rm dec}(\omega) }={R}(\omega,x_0) \,.
\end{equation}
Finally, the two-point function of the ECO is given by
\begin{align}
\mathcal{R}_{\rm ECO}(\omega) &\,= \frac{\alpha_{\rm ECO}(\omega)}{\beta_{\rm ECO}(\omega)}  = \mathcal{R}_{\rm BH}(\omega) + {\cal K}(\omega,x_0)\qty[B_{\rm in}^{\rm grow}(\omega)- {{R}_{\rm BH}(\omega)}B_{\rm out}^{\rm grow}(\omega)],
\label{eq: correlator general}
\end{align}
where
\begin{align}
\mathcal{K}(\omega, x_0) = \frac{{T}_{\rm BH}(\omega) {R}(\omega,x_0)}{1- {R}_{\rm BH}(\omega) {R}(\omega,x_0)}\, .
\end{align}
Using (\ref{eq:AdScc-cond}), (\ref{eq:AdSRTrel}), and (\ref{eq:AdScorrT}), we can further simplify (\ref{eq: correlator general}) to
\be \label{eq:corrAdSsimpl} \mathcal{R}_{\rm ECO}(\omega) \,= \text{Re}(\mathcal{R}_{\rm BH}(\omega)) + i\left({1 + R_{\rm BH}R \over 1- R_{\rm BH}R} \right) \text{Im}(\mathcal{R}_{\rm BH}(\omega)).\ee
At this point a natural question arises: How do we compare (\ref{eq:corrAdSsimpl}) to hybrid WKB expression for the propagator  (\ref{eq:RWKB})? We address this question in the next section.

\subsection{Comparison to Hybrid WKB}\label{sec:compareWKB}
AdS asymptotics differ from flat asymptotics in that the scalar wave does not ``leak'' out at infinity. This means that a scalar wave in a spacetime without a horizon (or other absorbing boundary conditions) only admits normal (real) modes instead of quasinormal modes (that have a finite imaginary part). The scalar wave will explore the inner structure of the ECO (which can be arbitrarily complicated, and take arbitrarily long), but will always end up being reflected out.

Since there is no dissipation and we have only normal modes $\omega_n$, we can assume the correlator $\mathcal{R}_{\rm ECO}(\omega)$ is real for real $\omega$. From (\ref{eq:corrAdSsimpl}) and using (\ref{eq:AdSRTrel}), it is easy to see that this is equivalent to the condition $|R| = 1$. It is then convenient to define the black hole phase $e^{i\theta_{\rm BH}}$ and the ECO phase $\theta_{\rm ECO}$ as:
\be R_{\rm BH} \equiv e^{i\theta_{\rm BH}}, \qquad  R \equiv e^{i\theta_{\rm ECO}},\ee
as then (\ref{eq:corrAdSsimpl}) becomes simply:
\be \label{eq:AdScorrR1} \mathcal{R}_{\rm ECO}(\omega) \,= \text{Re}(\mathcal{R}_{\rm BH}(\omega)) - \cot \left(\frac{\theta_{\rm BH} + \theta_{\rm ECO}}{2}\right) \text{Im}(\mathcal{R}_{\rm BH}(\omega)).\ee

We can further assume that the potential $V(x)$ has two turning points $x_\pm$ where $V(x_\pm)=0$ --- for example, this is the case for the wormhole we discuss in section \ref{sec:AdS3WKB:ext},\footnote{Note that both the wormhole of section \ref{sec:AdS3WKB:ext} as well as the microstate geometries in \cite{Bena:2019azk} have extremally rotating BTZ asymptotics, whereas in section \ref{sec:wavesAdS} and \ref{sec:compareWKB} we are only considering non-rotating AdS asymptotics. The hybrid WKB analysis can also be performed for the non-rotating Solodukhin wormhole, with approximate result $\Theta \approx |\omega| L_\lambda$ \cite{nextpaper}.} and for the actual microstate geometries discussed in \cite{Bena:2019azk}. Then, comparing (\ref{eq:AdScorrR1}) to the propagator (\ref{eq:RWKB}) of the hybrid WKB method propagator as in section \ref{sec:AdS3WKB:ext} or \cite{Bena:2019azk}, we arrive at the relation:
\be \theta_{\rm ECO} = \pi-\theta_{\rm BH} + 2\Theta = \pi-\theta_{\rm BH} + 2 \int^{x_+}_{x_-} \sqrt{\abs{V(z)}} \dd{z}.\ee
If we model the behaviour of the ECO as a ``brick wall'' with total reflection at a radius $x=x_0$, we take $R(\omega,x_0) = -e^{-2i\omega x_0}$ so that $\theta_{\rm ECO} = \pi-2\omega x_0$. Moreover, for the non-rotating Solodukhin wormhole (in analogy with (\ref{eq:thetaextwh})), we would have approximately $\Theta \approx \omega L_\lambda$, so that:
\be x_0 = -L_\lambda  + \frac{\theta_{\rm BH}}{2\omega},\ee
For $\omega$ and $L_\lambda$ large, the second term goes as $\mathcal{O}(\omega^0, L_\lambda^0)$ for the non-rotating BTZ black hole, and so $x_0\approx -L_\lambda$ is the position at which we would need to put the totally reflective wall in order to model the behaviour of the Solodukhin wormhole.

\section{Discussion}\label{sec:discuss}

We end with two observations that follow from combining our general analysis in our toy model wormhole geometries with recent developments in supersymmetric black hole microstate geometries. First, by using the travel time down a microstate throat \cite{Bena:2019azk}, we argue that the origin of quantum corrections to the black hole can be interpreted as sitting exponentially close to the horizon, both in terms of the Planck length and in terms of the black hole entropy. Second, we can perform an ensemble average over wormhole geometries that reveals how the echo amplitude generically gets exponentially suppressed in a typical state, and how it follows that the Poincar\'e recurrence time is doubly-exponential in the entropy.

In order to make generic statements that go beyond supersymmetric/extremal black holes, for simplicity we will use the non-extremal, non-rotating wormhole solution of section \ref{sec:DSWHnonrot} as a basis for those statements. We explain below where appropriate care has to be taken when comparing to current supersymmetric microstate literature.

\subsection{Where are the Quantum Corrections?}\label{sec:qucorr}
If we imagine or model a black hole microstate (geometry) as (quantum) corrections around the black hole horizon, one can then wonder \emph{where} these quantum corrections are located with respect to the horizon. Is the geometry made up out of seeming perturbative corrections within a Planck distance from the horizon? For the wormhole toy model \eqref{eq:WH3metric}, this would mean that $\lambda \propto \ell_P$, with $\ell_P$ the Planck length. Or, alternatively, are the corrections exponentially suppressed near the horizon, such that $\lambda \propto \exp(-L/\ell_P)$ for some reference length $L$? We could equivalently formulate this distinction as corrections that are perturbative or non-perturbative in the black hole entropy $S$. Here, we will first make a connection to recent 4D gravity literature and then answer this question by combining our wormhole analysis with results from recent microstate literature.

\subsubsection{Intuition in four-dimensional asymptotically flat space}

To quantify how compact objects (such as microstate geometries) differ from a black hole, one can define a `closeness parameter'
$\varepsilon$, as suggested in \cite{Cardoso:2019rvt}. For example, for a very crude model of a spherically symmetric reflective surface at a location $r_0$, a natural choice for this parameter is given by $r_0 = r_S(1 + \varepsilon)$ with $r_S$ the Schwarzschild radius $r_S = 2M$. One recovers the black hole metric in the limit $\varepsilon \to 0$. For more sophisticated objects than simple reflecting surfaces, one can often still determine a location of $r_0$ of the ``horizon structure'' or corrections based on natural criteria, such as a location where the Kretschmann scalar is noticably different,
or the radius beyond which most of the mass is concentrated (for boson star configurations), or some other criteria of distance or location. 
For reference, we can rewrite the definition of the closeness parameter $\varepsilon$ for a Schwarzschild black hole as:
\begin{equation}
\varepsilon \approx 1.4 \times 10^{-39}\left(\frac {r_0-r_S}{\ell_P} \right)\left(\frac{M_\odot}{M}\right)\,.
\end{equation}
We see that $\varepsilon \sim 10^{-39}$ corresponds to a coordinate distance of a Planck length from the horizon for a solar mass black hole.
For our wormholes, the role of closeness parameter is essentially played by $\lambda$.

By quantifying models in terms of $\varepsilon$, we can introduce a notion of ``distance" of the structure to the event horizon. Other notions of distance can also be considered instead:
\begin{itemize}
\item The parameter $\varepsilon = (r_0-r_S)/r_S$, the relative coordinate distance of the corrections.
\item The proper distance $\Delta s$ of the corrections from the horizon.
\item The (minimal) value of the redshift factor $\sqrt{|g_{tt}|}$, at the location of the corrections.
\item The travel time $\Delta t$ as seen by an asymptotic observer of a light ray from, say, the light ring to the location of the corrections.
\end{itemize}
For simple structures that are being compared to the four-dimensional Schwarzschild metric, we can easily see that these measures are all equivalent, as pointed out in \cite{Cardoso:2019rvt}. The proper distance is related to $\varepsilon$ as:
\begin{equation}
\Delta s \equiv \int_{r_S}^{r_S(1+\varepsilon)} \sqrt{g_{rr}}dr \approx 2\sqrt{\varepsilon} r_S,\quad \text{or} \quad 
\varepsilon \approx 2.1 \times 10^{-78}\left(\frac {\Delta s}{\ell_P}\right) \left(\frac{M_\odot}{M}\right)\,.
\end{equation}
The redshift that corresponds to it is
\begin{equation}
\sqrt{|g_{tt}|} \approx \varepsilon^{1/2}
\end{equation}
and the travel time from the light ring ($r = 3r_S/2$) to the structure is
\begin{equation}
\Delta t = M (1 - 2 \varepsilon - \log (4\varepsilon^2)) \approx -2 M \log \varepsilon.
\end{equation}

However, for a wormhole, or a microstate geometry, some of the above notions may not be related or even well-defined, and so cannot be used to define or determine $r_0$. First of all, such an object has no interior, and one can only make observations from the outside. This makes it impossible to have a simple notion of coordinate distance $r_0$ or proper distance from the would-be horizon. 
Second, microstate geometries are known to have ergoregions, where the redshift factor $\sqrt{g_{tt}}$ vanishes. Rotating black holes also have ergoregions; however, the ergoregions of the microstates are located at different locations than the black hole ergoregions, so keeping minimal redshift as the notion of the location of the structure could easily give the wrong idea.

For such geometries, then, we will use the travel time as measured by an asymptotic observer as a well-defined notion to quantify the deviation from the black hole metric.  Such an approach has partly been followed already in \cite{Bena:2006kb} where the travel time was directly related to the mass gap (the energy of the lowest excitation in the throat) and is thus directly related to redshift at the `cap', where the black hole-like throat smoothly ends.

\subsubsection{Perturbative or non-perturbative?}

We follow an approach based on the equivalences mentioned for four-dimensional solutions. In particular, we compare the travel time in wormholes and microstate geometries with the time in the corresponding BTZ metric to travel from the asymptotic boundary to a location $r_0$ and back. This allows us to determine, in terms of the wormhole or microstate geometry parameters, the fictional radius $r_0$ where we would naively have to put reflecting boundary conditions in the black hole geometry to mimick the microstructure --- and hence, give an indication of the ``location'' of quantum corrections to the BTZ geometry as coming from our wormhole or microstate geometries. 

There are two caveats we wish to highlight with this approach if one wishes to apply it more generally. First, getting the travel time from a wave equation or geodesic equation is technically possible for a simple, integrable model such as our wormhole geometries, or the separable superstrata solutions considered in \cite{Bena:2019azk}. However, a more typical microstate geometry will not have a separable wave equation or  geodesic equation. Due to mixing between radial and angular modes, it becomes harder to find or calculate a unique, well-defined infalling travel time. Thus, the location of the would-be radius $r_0$ one could derive in this setup would be at best an upper bound. The second caveat is that the travel time (for example, as measured by the time between echoes) is only a crude measure of microstructure and might miss other interesting structures in the black hole throat, such as the tidal stresses or curvature effects that have been advocated to play a role at smaller timescales than the travel time, i.e. $t_{\rm tidal} \ll \Delta t$, see for example \cite{Tyukov:2017uig,Bena:2019azk,Bena:2020iyw}.


\paragraph{Relating $\lambda$ to $r_0$ for wormholes}
First, we can discuss the Solodukhin wormhole as an example of where we can apply the above reasoning; this was already partly analyzed in \cite{solodukhin_restoring_2005}. We focus here on the general non-extremal BTZ black hole and its corresponding wormhole for generality, taking $r_+ > r_- \geq 0$.

For a radially infalling null geodesic in the BTZ metric, the travel time from the boundary to a location $r_0$ and back to the boundary is given by:\footnote{Radially infalling means we take $L_\phi = \frac {d\phi}{d\tilde\lambda} + N_\phi \frac{dt}{d\tilde\lambda} = 0$, where $\tilde\lambda$ is an affine parameter along the geodesic. The travel time is then defined as $\Delta t = 2\int_{r_0}^\infty \frac{dt}{dr} dr = 2 \int X^{-1}dr$. The proper distance from the horizon is defined as $\Delta s = \int _{r_+}^{r_0} X^{-1/2} dr$.}
\begin{equation}
\Delta t_{\rm BTZ} = \frac{2R^2}{r_+^2-r_-^2} \left(r_+ \tanh ^{-1}\left(\frac{r_+}{r_0}\right)-r_- \tanh ^{-1}\left(\frac{r_-}{r_0}\right)\right).\label{eq:BTZ_traveltime}
\end{equation}
When $\varepsilon = \frac{r_0-r_+}{r_+}\ll 1$, this becomes:
\begin{equation}
\Delta t_{\rm BTZ}  = \frac{r_+ R^2}{r_+^2-r_-^2} \log \frac{2}{\varepsilon}+ O(\varepsilon^0)\,.
\end{equation}
We need to compare this to the  travel time in our wormhole toy model. The time it takes for a signal to travel from the boundary through the wormhole and come back to the boundary is (to leading order in $\lambda$) twice the throat length: $\Delta t_{\rm WH} =  2 L_\lambda$. 
An analogous computation of the travel time for the non-extremal rotating wormhole of sections \ \ref{sec:DSWHnonrot} and \ref{sec:extDSWH} \footnote{To generalize the non-rotating wormhole metric (\ref{eq:WH3metric}) to the rotating, non-extremal case, we can simply change $X(r)$ by taking $r_+\rightarrow r_\lambda$ in the BTZ metric (\ref{eq:BTZmetric}), while leaving $Y(r)$ and $N_\varphi(r)$ unchanged.} gives: 
\begin{equation}\label{eq:Llambdar-}
L_\lambda = \frac{r_+ R^2}{r_+^2-r_-^2} \log \frac{16}{\lambda^2}.
\end{equation}
Then, equating $\Delta t_{\rm BTZ} = 2 L_\lambda$, we find a relation between $\lambda$ and $\varepsilon$:
\begin{equation}
\varepsilon = \frac{\lambda^4}{128}~.
\end{equation}

For a more coordinate invariant notion of distance, one can evaluate the proper distance $\Delta s$ between $r=r_+(1+\varepsilon)$ and the horizon.
This quantity is related to $\varepsilon$ as
\begin{equation}
\Delta s = \frac{r_+ R}{2}\sqrt{\frac{2 \varepsilon}{r_+^2-r_-^2}}+O(\varepsilon)~.
\end{equation}
The proper distance to the horizon diverges for an extremal wormhole, so we can only compare $\lambda$ to the radial location $r_0$ in the BTZ metric.
For the extremal BTZ black hole, we find that the travel time is given by:
\begin{equation}
\Delta t_{\rm BTZ} = \frac{R^2}{2r_+\varepsilon} +O(\varepsilon^0)\,.
\end{equation}
This is now no longer logarithmic, but inversely proportional to the (coordinate) distance to the horizon. The travel time in the extremal wormhole is again given by $\Delta t_{\rm WH}=2 L_\lambda$, with $L_\lambda= 2\frac{R^2}{r_+}\frac{1}{\lambda^2}$ from eq.\ (\ref{eq:WH3extdefthroatlength}). Setting $\Delta t_{\rm BTZ}=2 L_\lambda$ again then gives to leading order:
\begin{equation}
\epsilon = \frac{\lambda^2}{8}\,.
\end{equation}
Therefore, both for the non-extremal and the extremal wormholes, we see that the  parameter $\lambda$ can be interpreted as a measure for the (proper) distance that the wormhole throat sits from the would-be horizon radius.

\paragraph{Relating distance to entropy.}
As we mentioned above, an important question to ask is: at what distance from the (would-be) horizon should we expect microstructure to arise? For the wormholes we consider, the two most obvious options for the location of the structure (i.e. the wormhole throat) are either perturbative or non-perturbative corrections.

By perturbative corrections, we mean a wormhole parameter $\lambda$ (and thus distance to the horizon) that scales as $\lambda \sim G_N$, while non-perturbative is taken to be exponential in the gravitational coupling. Through the Bekenstein-Hawking entropy of the BTZ black hole,
\begin{equation}
S_{\rm BH} = \frac{\pi r_+}{2G_N}\label{eq:entropyBTZ}
\end{equation}
we can also interpret perturbative corrections as scaling with the black hole entropy as $\lambda_{\rm pert} \propto 1/S_{\rm BH}$, while non-perturbative corrections have $\lambda_{\rm non-pert} \propto \exp(-\alpha S_{\rm BH})$ for some dimensionless constant $\alpha$.

Using holography, one could also consider the CFT central charge $c$ instead of $G_N$. By the Brown-Henneaux central charge relation
\be
c = \frac{3 R}{2 G_N} \label{eq:BHcentralcharge}
\ee
we could also have defined perturbative corrections as $\lambda \sim 1/c$, while non-perturbative corrections are $\lambda \sim \exp(-a c)$ for some constant $a$. These relations are explored in \cite{solodukhin_restoring_2005}. These two viewpoints are related through the Cardy formula for the CFT entropy, $S_{\rm BH} = 2\pi \sqrt{c\,E/6}$, where $E = (c/6) (r_+/R)^2$ is the energy of the state in the CFT. We will choose to work with the entropy as opposed to the central charge as this interpretation does not necessarily hinge on the existence of a CFT dual.

\paragraph{Extra input from microstate geometries}
For a wormhole, we cannot directly determine the relation of $\lambda$ to $G_N$ or the entropy. Instead, we will use an indirect route of comparing our wormhole toy model to black hole microstate geometries. Recently, the travel time for a scalar wave packet in superstrata was considered in \cite{Bena:2019azk}. The travel time from the boundary to the cap is determined by the mass gap $E_{\rm gap} = (N_1 N_5)^{-1}$ \cite{Tyukov:2017uig,Bena:2007qc,Bena:2006kb}, the energy of the lowest energy excitation down the throat of the supersymmetric solution:
\begin{equation}
\label{eq:deltatms}\Delta t_{\rm microstate} \approx E_{\rm gap}^{-1} R_y = {N_1 N_5}R_y\,.
\end{equation}
For our extremal wormhole, we found $\Delta t = 2 L_\lambda = \frac{4 R^2}{r_+ \lambda^2}$.
Equating these two travel times, we find
\begin{equation}
\lambda^2 \propto \frac{1}{\sqrt{N_1 N_5 N_P}}\propto \frac{1}{S_{\rm BH}}~,
\end{equation}
where we used $R=(Q_1 Q_5)^{1/4}$, $r_+ = \sqrt{Q_p}$, $\frac{Q_1 Q_5}{Q_P} = \frac{N_1 N_5}{N_P} R_y^2$, and the entropy of the BMPV black hole with the same charges of the microstate is $S_{\rm BH} = 2 \pi\sqrt{N_1 N_5 N_P}$.\footnote{We used that the $(1,0,n)$ superstratum with parameter $b \gg a$ has nearly-vanishing quantized angular momentum $j_L$, such that the entropy is given to excellent approximation by $S_{\rm BH} = 2\pi \sqrt{N_1N_5N_P- j_L^2} \approx 2\pi \sqrt{N_1 N_5 N_P}$, see \cite{Heidmann:2019xrd, Bena:2019azk}.}

Imagine now that we add a small amount of non-extremality to the microstate, for instance by adding a small number of anti-branes down the D1-D5-P throat (or, equivalently, right-movers in the CFT dual to the BTZ solution).
The black hole then becomes near-extremal, and the behavior of travel time becomes logarithmic in $\lambda$.
However, we do not expect such a small change to the superstratum geometry to drastically change the travel time, so we assume it is still proportional to the mass gap $E_{\rm gap}$\footnote{Note that $E_{\rm gap}$ is still meant to be the SUSY mass gap, and should not be confused with the mass gap of excitations above the near-extremal solution, which is expected to scale as $e^{-S_{\rm BH}}$.
This introduces a new scale that should be decoupled from the travel time itself, as it is related to jumps in energy to \emph{different} microstates, see \cite{Bena:2018bbd}.}
Then, assuming that at leading order the non-extremality does not alter the relation \eqref{eq:deltatms}, nor the relations between $R$, $r_+$ and the charges $N_1$, $N_5$, $N_P$, we find that equating $\Delta t_{\rm microstate} = 2 L_\lambda$ (see eq.\ \eqref{eq:Llambdar-}) leads to
\begin{equation}\label{eq:loglambdaS}
\log \frac{16}{\lambda^2} \propto \left(1-\frac{r_-^2}{r_+^2}\right)\sqrt{N_1 N_5 N_P} \propto S~,
\end{equation}
Hence, we have that:
\be \label{eq:lambdaexpS}
\lambda \propto e^{-\alpha S_{\rm BH}}
\ee
This exponential suppression of $\lambda$ suggests that the `closeness parameter' $\varepsilon\propto \lambda^4$ can be interpreted as being sub-planckian, even though the microstate geometry itself is well-behaved in the supergravity limit. However, note that this prefactor $\alpha \sim 1-r_-^2/r_+^2$, which is in principle small near extremality. One can conjecture that (\ref{eq:lambdaexpS}) still remains valid further away from extremality (with $\alpha\sim \mathcal{O}(1)$); it would be interesting to investigate whether this is indeed the case in future work.

\subsection{Does a Typical Black Hole Microstate Have Gravitational Wave Echoes?}\label{sec:lambdaensemble}
Following Betteridge's law, the answer to the question in the title is: most likely, no.

There have been many studies of black hole mimickers giving rise to discernible echo structure in the post-ringdown gravitational wave signal, coming from the partial or complete reflection of the wave around the would-be horizon scale --- see for example \cite{Cardoso:2016rao,Cardoso:2016oxy,Cardoso:2017cqb,Cardoso:2017njb,Barack:2018yly}; there have even been serious attempts to find echoes in existing gravitational wave signals \cite{Abedi:2016hgu,Ashton:2016xff,Abedi:2017isz,Abedi:2018pst,Abedi:2018npz,Abedi:2020sgg}. 

As we discussed above in section \ref{sec:qucorr}, the (quantum) corrections to the (non-extremal) geometry are expected to be exponentially suppressed as  $e^{-\alpha S_{\rm BH}}$, so one might immediately conclude that such corrections can never be measureable. How, then, could we expect to see these echoes?

The answer is simple: in the echo time, we pick up a logarithm, such that $\Delta t \propto \log \lambda \propto 1/S_{\rm BH}$ and corrections to the travel time are perturbative. Thus, there is a clear difference between geometric corrections and the corrections to observables such as echoes in wave response.

However, if we are to interpret the black hole mimickers (such as the Solodukhin wormhole studied here) as serious candidates or toy models of black hole microstates, then a \emph{typical} black hole microstate would be an arbitrary superposition of such states. One such state, described by one particular value of $\lambda$, indeed has large corrections and an observable echo signal. But when forming a typical state by superposition of many such states, one gets exponential suppression of the \emph{amplitude} of subsequent echoes by interference.

As we discussed in sections \ref{sec:wavesAdS} and \ref{sec:compareWKB}, echoes in an asymptotically AdS spacetime have a direct counterpart in the two-point correlator of the dual holographic CFT, where many results are known that describe in detail the expectation for the behaviour of such correlators.  Generically, one expects to find that the behaviour of this correlator in a thermal unitary quantum field theory has the following features \cite{Maldacena_2003,Barbon:2003aq}:
\begin{itemize}
\item at early times, the correlator decays exponential as $\ev{\mathcal{O}\mathcal{O}} \sim \exp\qty(-2\pi{ 2\Delta\over \beta}t)$ (where $\Delta$ is the scaling dimension of the operator $\mathcal{O}$);
\item after the initial falloff, there are random fluctuations; these are suppressed by the entropy as $\ev{\mathcal{O}\mathcal{O}}\lesssim e^{-S/2}$;
\item at very long time scales there is a (Poincar\'e) resurgence expected, at a time that scales doubly exponential with the entropy.
\end{itemize}
Can we recreate this behaviour by performing a statistical average of microstates with supergravity duals?
For realistic microstates of the BMPV black hole, performing such an averaging is beyond current technology; only certain atypical states have known supergravity duals.  However, using the Solodukhin wormhole as a toy model for microstates, we can consider an ensemble of wormholes with different $\lambda$ parameters, and give some crude estimates for the generic expectations of the correlator in a typical state.



\begin{figure}[!ht]
\begin{subfigure}{.5\textwidth}
  \centering
  \includegraphics[width=0.95\linewidth]{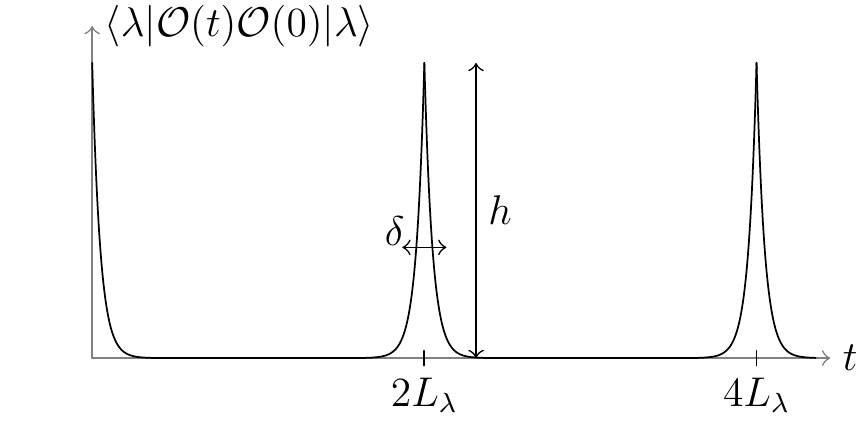}  
  \caption{Single $\lambda$ geometry}
  \label{fig: one lambda}
\end{subfigure}
\begin{subfigure}{.5\textwidth}
  \centering
  \includegraphics[width=0.95\linewidth]{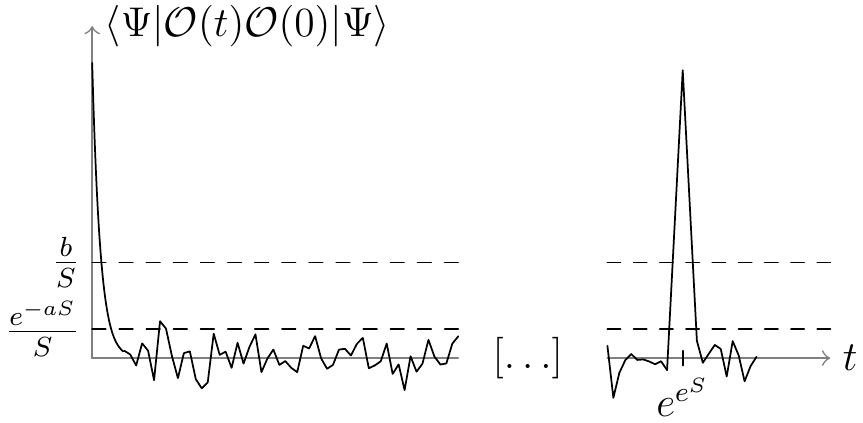}  
  \caption{Random superposition of $\lambda$ geometries}
  \label{fig: ensamble of lambdas}
\end{subfigure}
\caption{Rough sketch of correlator in a single wormhole geometry $|\lambda\rangle$, and in a typical state $|\psi\rangle$, which is a random superposition over many such wormhole geometries}
\label{fig: correlators}
\end{figure}

\paragraph{A microcanonical ensemble}
Denote a single Solodukhin wormhole state as $| \lambda\rangle$ with throat length $L_\lambda = -2r_0 \log \lambda$, and assume that for any two such states we have $\langle \lambda'|\lambda\rangle=\delta_{\lambda,\lambda'}$.\footnote{If the states $|\lambda\rangle$ are more accurately described as coherent states (which would be natural for classical supergravity solutions), then their overlap would be exponentially small but not zero.} We will be interested in two-point functions of a (light) scalar operator $\Phi$ in these states; we assume $\Phi$ is light enough so that it does not deform the (heavy) state $|\lambda\rangle$ much, meaning also $\langle \lambda' | \Phi(t)\Phi(0) | \lambda \rangle \approx 0$. Following the general structure of the two-point functions we have found (see section \ref{sec:WHPSprop}, especially (\ref{eq:naiveWHGR})), let us assume here a schematic two-point function of the form:
\be \langle \lambda | \Phi(t)\Phi(0) | \lambda \rangle = G_0(t) + \sum_{n=0}^{\infty} G_1(t-2n L_\lambda), \label{eq: two point}\ee
where $G^\text{BH}(t) = G_0(t) + G_1(t)$ is the corresponding black hole correlator, which peaks at $t=0$ and then quickly decays (see the bullet points above). The $n>1$ terms are those that give the ``echoes'' at late times, starting with the first echo at $t\approx 2L_\lambda$.

Now, we assume the black hole can be described by a microcanonical ensemble of such wormhole states $|\lambda\rangle$ with $\lambda_{min}\leq \lambda \leq \lambda_{max}$ and corresponding throat lengths $L_{\lambda,max},L_{\lambda,min}$. We assume $\lambda_{min}\ll \lambda_{max}\ll 1$ (and so $r_0\ll L_{\lambda,max}\ll L_{\lambda,min}$). The number of states that spans this ensemble is $\#(\{|\lambda\rangle\})=e^S$, where $S$ is the black hole entropy.
A generic state in the microcanonical ensemble is given by:
\be |\Psi\rangle = \sum_{\{|\lambda\rangle\}} b_\lambda |\lambda\rangle,\ee
where $b_\lambda$ are randomly selected\footnote{More precisely: chosen with the Haar measure.} complex numbers such that $\sum_{\{|\lambda\rangle\}} |b_\lambda|^2=1$. This implies that on average $|b_\lambda|^2\sim e^{-S}$. Then our assumptions imply:
\be \langle \Psi | \Phi(t)\Phi(0) | \Psi \rangle = \sum_{\{|\lambda\rangle\}} |b_\lambda|^2 \left(G_0(t) + \sum_{n=0}^{\infty} G_1(t-2n L_\lambda)\right) = G^\text{BH}(t) + \tilde{G}^{mc}_1(t),\ee
where we have defined $\tilde{G}^{mc}_1(t)$, the deviation from the black hole correlator.
We are interested in the behaviour of $\tilde{G}^{mc}_1(t)$ at late times. Let us model $G_1$ by a simple theta function of height $h$ and width $\delta$, which captures the fact that the correlator $G^\text{BH}(t)$ is peaked around $t=0$ and small for late times, see figure \ref{fig: one lambda}:
\be G_1(t) = h[ \Theta(t-\delta) - \Theta(t+\delta)],\ee
so that:
\begin{align}
\label{eq:Gmcline1} \tilde{G}^{mc}_1(t) &= \sum_{\{|\lambda\rangle\},n>0} |b_\lambda|^2   h \left[ \Theta(t-2nL_\lambda-\delta) - \Theta(t-2nL_\lambda+\delta)\right]\\
 \label{eq:Gmcline2}&\approx h e^{-S}\sum_{n>0}\left(\lambda_{1,n}-\lambda_{0,n}\right) = h e^{-S}\sum_{n>0}\left( \exp\left( -\frac{t-\delta}{4nr_0}\right)- \exp\left( -\frac{t+\delta}{4nr_0}\right)  \right)
 \end{align}
 where we have used $|b_\lambda|^2\sim e^{-S}$ and $\lambda_{1,n}$ (resp. $\lambda_{0,n}$) are the largest (resp. smallest) possible $\lambda$ for which the theta functions in (\ref{eq:Gmcline1}) don't vanish. The sum over $n$ has a finite range as we must have that $\lambda_{min}\leq\lambda_{0,n} < \lambda_{1,n}\leq\lambda_{max}$. This gives:
 \be \frac{t+\delta}{2L_{\lambda,min}} \leq n \leq \frac{t-\delta}{2L_{\lambda,max}}.\ee
 We now approximate the sum over $n$ in (\ref{eq:Gmcline2}) by an integral and use $t\gg \delta$ and $L_{\lambda,min/max}\gg r_0$ to find:
 \be \label{eq:Gmcsmall} \tilde{G}^{mc}_1(t) \approx h\delta e^{-S}\left( \frac{e^{-L_{\lambda,max}/(2r_0)}}{L_{\lambda,max}} - \frac{e^{-L_{\lambda,min}/(2r_0)}}{L_{\lambda,min}}\right) \approx  h\delta e^{-S}\frac{\lambda_{max}}{L_{\lambda,max}}    ,\ee
where we used $L_{\lambda,min}\gg L_{\lambda,max}$ and $\lambda_{max}\gg \lambda_{min}$.
Equation \eqref{eq:Gmcsmall} shows that the correlator is suppressed by $e^{-S}$, $\lambda_{max}$ and $L_{\lambda,max}^{-1}$, such that $\tilde{G}^{mc}_1(t)/\tilde{G}^{mc}_1(0)\ll 1$.
The conclusion is that the difference of the two-point correlator in the typical state $|\Psi\rangle$ from the black hole correlator is negligibly small.

To determine more precisely the suppression of the correlator in terms of the entropy, one should keep in mind that $S$ and $\lambda_{max}$ are related in some way that we have yet to specify in our crude model.
For example, if the states $\{|\lambda\rangle\}$ differ in $\lambda$ by some constant spacing $\Delta\lambda$, we find $e^{S} = \frac{\lambda_{max}-\lambda_{min}}{\Delta \lambda}\approx \frac{\lambda_{max}}{\Delta\lambda}$.
The exponential factors in the entropy then cancel and our correlator is only polynomially suppressed in the entropy: $\tilde{G}^{mc}_1(t)/\tilde{G}^{mc}_1(0) \propto b/S$, for some positive constant $b$. By contrast, if we assume the scaling of $\lambda$ with the entropy $S$ in equation \eqref{eq:lambdaexpS}, then  the suppression is $e^{-a S}/S$ for some positive constant $a$, see figure \ref{fig: ensamble of lambdas}.

This simple crude calculation in our toy model has shown us the generic expectation for typical black hole microstates: any ``echo structure'' that might be found in individual, highly non-typical microstates (here, in any single state $|\lambda\rangle$), is generically washed away in a typical state that is a random superposition of these states.

\paragraph{Recurrence time}
Instead of studying  $\tilde{G}^{mc}_1(t)$ at a generic (late) time as we have done above, we can ask the reverse question: at what time $t_R$ do we expect to find $\tilde G^{mc}_1(t_R)\approx h$, i.e. having an amplitude peak similar to the original (black hole) peak at $t=0$. To have such a recurrence at a time $t_R$, we need that approximately \emph{all} the states in the ensemble, $\lambda_i\in \{|\lambda\rangle\}$, satisfy:
\be t_R - 2n_i L_{\lambda,i} - \delta \leq 0 \leq t_R - 2 n_i L_{\lambda,i} + \delta,\ee
for some collection of integers $\{ n_i \}$. This means that, for every pair $i,j$, we need:
\be \left| n_i \frac{L_{\lambda,i}}{\delta} - n_j \frac{L_{\lambda,j}}{\delta}\right| \leq 1.\ee
Unless there is a special way the ensemble $\{|\lambda\rangle\}$ is constructed, the $\lambda_i$'s and thus the $L_{\lambda_i}$'s will generically be random; together with $\delta \ll L_{\lambda,max}$ this implies the solution must be (multiples of):
\be n_i \approx \prod_{j\neq i} \frac{L_{\lambda,j}}{\delta}.\ee
Then:
\be t_R \approx  2 n_i L_{\lambda,i}  \approx  2\delta \prod_i \frac{L_{\lambda,i}}{\delta}. \label{eq: recurrence}\ee
If we take $L_{\lambda,max}= N \delta  $ (with $N\gg 1$), then we can give a bound on $t_R$ as:
\be t_R \geq (2\delta) N^{e^S} ,\ee
which gives the expected characteristic doubly exponential behaviour, see figure \ref{fig: ensamble of lambdas}.

To conclude, we have seen that even our very crude toy model of averaging over microstate geometries, the correlator behaviour matches remarkably well with the generic QFT expectations we discussed above:
\begin{itemize}
\item At early times, the decay is exactly the same as the one for a black hole, since the $n=0$ contribution of (\ref{eq: two point}) is precisely the black hole correlator, and is independent of the particular heavy state $\ket{\lambda}$).
\item After the initial decay, there are ``random'' fluctuations due to the individual microstate in question, given by (\ref{eq:Gmcsmall}). The  smallness of these fluctuations depends on the specifics of the distribution of $\lambda$'s within the ensemble, and so depends on the details of our toy model.
\item For an ensemble of appropriately random $\ket{\lambda}$'s, the recurrence time (\ref{eq: recurrence}) is indeed doubly exponential in the entropy.
\end{itemize}
In a more realistic model, such as superstratum microstate geometries, one would have more parameters to consider (other than just $\lambda$ as in the wormhole). Moreover, the echoes for most microstates will not necessarily be equally spaced or clearly discernable between microstates. Also, it is not obvious that the contribution from off-diagonal terms in the correlator is negligible --- here, we simply assumed $\langle \lambda'|\lambda\rangle\approx 0$. These are all interesting open questions that we would like to return to in future work.

%

\section*{Acknowledgments}
We thank Nikolay Bobev, Geoffrey Comp\`ere, Christian Maes for useful discussions, and especially Pierre Heidmann and Ruben Monten for explanations about \cite{Bena:2019azk} and related topics. BV would also like to thank the participants and organizers of the (virtual) Black Hole Microstructure Conference at the IPhT, CEA Saclay, June 8-12, 2020, for useful comments on his talk on this paper. The work of DRM is supported by the ERC Starting Grant 679278 Emergent-BH.  The work of VD is supported by BIJZONDER ONDERZOEKSFONDS 2016 (C16/16/005) and a doctoral fellowship from the Research Foundation - Flanders (FWO). The work of VSM is supported by a doctoral fellowship from the FWO. TM and BV are partially supported by the National Science Foundation of Belgium (FWO) grant G.001.12, the European Research Council grant no. ERC-2013-CoG 616732 HoloQosmos, the KU Leuven C1 grant ZKD1118 C16/16/005, the FWO Odysseus grant G0H9318N and the COST actions CA16104 GWVerse and MP1210 The String Theory Universe.

\bibliographystyle{toine}
\bibliography{echoes}

\end{document}